\documentclass[aps,twocolumn,prb]{revtex4}
\usepackage[dvips]{graphicx}
\usepackage[dvips]{color}
\usepackage{amsmath, amssymb, bm}

\newcommand{\n}{{\vec{n}}}
\newcommand{\z}{{\bm z}}

\newcommand{\A}{\mathcal {A}}
\def\etal{~\textit{et~al.}} 
\newcommand{\ra}{\rangle} 
\newcommand{\la}{\langle} 

\newcommand{\up}{\uparrow}
\newcommand{\dn}{\downarrow}

\newcommand{\K}{{\cal K}}
\newcommand{\J}{{\cal J}}
\newcommand{\rhodual}{{\rho_{\rm dual}}}
\newcommand{\rhodualqmin}{{\rho_{\rm dual}(q_{\rm min})}}
\newcommand{\rhodualxL}{{\rho_{\rm dual} \cdot L}}

\begin{document}
\title{Comparative study of Higgs transition in one-component and
two-component lattice superconductor models}
\author{Olexei I. Motrunich}
\affiliation{Department of Physics, California Institute of Technology,
Pasadena, CA 91125}
\author{Ashvin Vishwanath}
\affiliation{Department of Physics, University of California,
Berkeley, CA 94720 \\
Materials Sciences Division, Lawrence Berkeley National
Laboratory, Berkeley, CA 94720}

\date{\today}

\begin{abstract}
{ Using Monte Carlo simulations, we study a Higgs transition in
several three-dimensional lattice realizations of the
noncompact CP$^1$ model (NCCP$^1$), a gauge theory with two complex
matter fields with SU(2) invariance. By comparing with a
one-component theory, which is well understood and has continuous
transition in the inverted XY universality class, we argue that the
two-component case also has continuous Higgs transition with a
larger correlation length exponent (i.e., it is ``more continuous''). 
The transition can become first order in the vicinity of a new 
``Molecular'' phase, which occurs in one of our models, 
but is continuous in a wide range of parameters away from this phase.
The situation is significantly clarified by studying a model
where the Molecular phase is entirely absent, and a wide regime
with a continuous transition can be readily established.
The two-component theory is also an effective description of the 
hedgehog-suppressed O(3) universality, and results are relevant for the 
recently discussed ``deconfined quantum criticality'' scenario for the 
continuous Valence Bond Solid to Neel quantum phase transition. }
\end{abstract}

\maketitle

\section{Introduction}
\label{sec:intro}

In this paper, we examine the nature of the Higgs transition in a
two-component theory with a dynamical gauge field in three
(classical) dimensions (3D).\cite{HLM} 
A schematic continuum action is
\begin{equation}
S = \int_{R^3} |({\bm \nabla} - i {\bm a}){\bm \Psi}|^2
+ m |{\bm \Psi}|^2 + u |{\bm \Psi}|^4
+ \kappa ({\bm \nabla} \times {\bm a})^2 ~.
\label{S_continuum}
\end{equation}
${\bm \Psi}$ has two complex components and ${\bm a}$ is the gauge
field that can take any real values (i.e., it is non-compact).
This theory is also called the noncompact CP$^1$ model
(NCCP$^1$).\cite{shortlight}  When the charged field ${\bm \Psi}$
is gapped, the gauge field is massless and there is a free photon,
but when ${\bm \Psi}$ condenses, ${\bm a}$ becomes massive -- this
is the Higgs transition. In Ref.~\onlinecite{shortlight}, we argued
that this action also describes an O(3) classical spin system with
prohibited hedgehog topological defects on long scales. 
In particular, the ordering transition in the hedgehog-suppressed model
and the above Higgs transition are in the same universality. The
nature of the transition is of interest in the context of the
recently discussed quantum phase transitions that lie outside the
Landau-Ginzburg-Wilson framework. \cite{deccp, XYring,
SU2ring1, SU2ring2, dimers3D} Thus, it has been proposed\cite{deccp,
XYring, SU2ring1, SU2ring2} that the Neel to Valence Bond Solid
transition in a spin-1/2 system with SU(2) symmetry can undergo a
continuous quantum phase transition described by the same
universality Eq.~(\ref{S_continuum}).

Here we present a detailed numerical analysis of several lattice
realizations of the theory~(\ref{S_continuum}) (we will refer to these as
lattice superconductor models\cite{DH}).  We extend the study of
Ref.~\onlinecite{shortlight} of the SU(2)-symmetric two-component model,
with the goal to check more thoroughly the proposed second-order nature
of the transition.
Here, we argue that our models have continuous Higgs transition over
wide parameter ranges, see Figs.~\ref{fig:phased}b,c. Note that in
one of our models, Fig.~\ref{fig:phased}b, we also find a parameter
range, in the vicinity of a third ``Molecular'' phase [produced when
gauge neutral pairs of the $\Psi_i$ particles of 
Eq.~(\ref{S_continuum}) condense], where the transition becomes first
order. This is always a possibility, and in the specific model
happens generically near this new phase.  We adopt two fresh
strategies to establish the nature of the transition in these
systems. First, in the regime where we believe the transition is
second-order, we compare with a one-component model for which it is
well established that the transition is continuous,\cite{DH} and
show that the transition in our two-component models is ``more
continuous'', in the sense that the thermodynamic singularities are
more weak.  Second, we introduce a lattice realization of the
model in Eq.~(\ref{S_continuum}), where the Molecular phase is
entirely absent. The Higgs transition is then shown to be continuous
over the entire phase boundary, which greatly simplifies the
interpretation. Our best estimates for the exponents at this
transition are $\nu = 0.7 - 0.75$ and $\eta = 0.2 - 0.4$. 
Recent Quantum Monte Carlo simulations of square lattice $S=1/2$ 
Heisenberg antiferromagnets with four spin exchange terms and 
SU(2) symmetry have also found a continuous relativistic transition 
between a Neel and Valence Bond Solid phase,\cite{SU2ring1, SU2ring2} 
which is believed to be in the same universality class as the 
transition studied here.  
The exponents obtained in those studies are $\eta=0.26(3)$ and
$\nu=0.78(3)$ in Ref.~\onlinecite{SU2ring1} and
$\eta=0.35(3),\,\nu=0.68(4)$ in Ref.~\onlinecite{SU2ring2}, consistent
with the ones reported here.

Other studies,\cite{Kuklov, Kragset, Chandrasekharan} particularly in 
two-component superconductor models where the SU(2) symmetry is broken 
down to U(1) (easy plane NCCP$^1$), have suggested that such Higgs 
transitions may always be first order.  
While those authors so far have found first order transitions in these 
U(1)$\times$U(1) models, we do not know any fundamental reasons why this 
needs to hold generally. 
This is particularly true in light of the evidence presented here for a 
continuous transition in the SU(2) symmetric model.  
Moreover, the studies in Refs.~\onlinecite{Kuklov, Kragset}
were performed on models where the Molecular phase is present, which
we suspect may be driving the first order nature of the transition.
In future work we will apply the ideas described here to obtain and
study models without a Molecular phase in this symmetry class as
well, which should help clarify the issue.

The study of the criticality in the theory~(\ref{S_continuum}) dates
back to Halperin, Lubensky, and Ma.\cite{HLM} At the fixed point
where the matter and gauge fields are decoupled ($e^2 \sim 1/\kappa
\to 0$), the gauge interaction is a relevant perturbation, and no
new stable fixed points were found in $d = 4-\epsilon$ treatment
when the number of components $N$ is less than 365.  This was
interpreted as an evidence for a fluctuation-induced first-order
transition.  For larger $N$, and also in a $1/N$ treatment directly
in $d=3$, the transition was found to be continuous. Later work by
Dasgupta and Halperin\cite{DH} showed that the transition in the
$N=1$ case can also be continuous.  The one-component superconductor
model is dual to the usual 3D XY model that has only short-range
interactions.  Concisely stated, one can fruitfully think about the
one-component system in terms of the Abrikosov-Nielsen-Olesen (ANO)
vortices which carry quantized gauge flux and have only short-range
interactions.  The duality claim then follows if we recall that the
3D XY model can be also viewed as a system of short-range
interacting current loops.  The Higgs phase where the ANO vortices
are gapped corresponds to the disordered phase of the XY model with
small loops, while the disordered phase of the charged
superconductor where the ANO vortices condense corresponds to the
ordered phase of the XY model with proliferated loops.  Thus, the
ANO vortices play a crucial role, which is difficult to capture in
the $4 - \epsilon$ treatment. We expect that the same physics
happens in the two-component case, but with the ANO vortices that
now have non-trivial internal structure. While our analytical
picture is not as complete as in the one-component case, since
we are unable to perform a duality transformation in the present
case, one of our models has a limit that strongly suggests such an
interpretation, and we confirm that the transition is continuous by
direct Monte Carlo studies on the lattice.

One of our main strategy exploring such new two-component matter -
gauge systems in Monte Carlo is to compare closely with the
well-understood one-component case.  We simulate both systems under
similar conditions and compare directly observables in the gauge
sector that detect the Meissner-Anderson-Higgs physics of charged
condensates. We also study separately new features in the
two-component system, such as the presence and possible condensation
of composite gauge-neutral fields.

The paper is organized as follows.  In Sec.~\ref{sec:models},
we define the models and describe their overall phase diagrams.
In Sec.~\ref{sec:transitions}, we focus on the Higgs transitions and
present several two-component cases where we argue the transition is
continuous.  We also discuss the observed systematics along the full
Photon to Higgs phase boundary and show that we can improve it by
suppressing possible other phases.
In Sec.~\ref{sec:concl}, we summarize the results and conclude
with an outlook.

\section{Models and their phase diagrams}
\label{sec:models}

We study the following three models.
The first model realizes a one-component lattice superconductor:
\begin{eqnarray}
S_{{\rm NCCP}^0} &=&
- {\cal J} \sum_{i, \mu} \cos(\theta_i - \theta_{i+\hat{\mu}} + a_{i\mu})
+ S_a ~,\\
S_a &=& \frac{\cal K}{2} \sum_\Box (\Delta \times a)^2 ~.
\label{nccp0}
\end{eqnarray}
A U(1) matter field $e^{i\theta}$ resides on the sites of a 3D
cubic lattice, while a non-compact gauge field $a_\mu$ resides on the
links; $S_a$ is the usual lattice Maxwell action for the gauge field.
We will use `NCCP$^0$' as a short-hand reference to this model, where
`CP$^0$' refers to one matter field component while `NC' emphasizes that
the gauge field is non-compact.\cite{shortlight, deccp}
This model and the ones below have two parameters $\J$ and $\K$.
The NCCP$^0$ model is our reference system for matter-gauge
simulations.

The second model realizes a two-component lattice superconductor:
\begin{equation}
S_{{\rm NCCP}^1, \text{model I}} = -\frac{\cal J}{2} \sum_{i,\mu}
\left( {\bm z}_i^\dagger {\bm z}_{i+\hat\mu} e^{i a_{i\mu}} + c.c.
\right)
+ S_a ~.
\label{nccp1}
\end{equation}
Here the matter field is a CP$^1$ field ${\bm z} = (z_\up, z_\dn)$,
$|{\bm z}| = 1$.
This model was introduced in Ref.~\onlinecite{shortlight} and studied
at $\K=0.6$ using small system sizes; the present work is an extension
of that study.

The third model is a different realization of the two-component case:
\begin{equation}
S_{{\rm NCCP}^1, \text{model II}} = S_{{\rm NCCP}^1, \text{model I}}
+ \sum_{i, \mu} \ln I_0 (\J |{\bm z}_i^\dagger {\bm z}_{i+\hat\mu}|) ~.
\label{nccp1wAF}
\end{equation}
Here $I_0(x)$ is the modified Bessel function.  Compared to the
model I, there is an additional antiferromagnetic interaction
between nearest-neighbor spins.  This interaction eliminates the
unwanted 'Molecular' phase from our phase diagram.  Our motivation
for this model will become clear soon, while here we only note that
the specific form allows some analytical understanding of the phase
diagram and that any such short-range interaction that respects the
symmetries of the problem does not change the universality of the
transition.

\begin{figure}
\centerline{\includegraphics[width=\columnwidth]{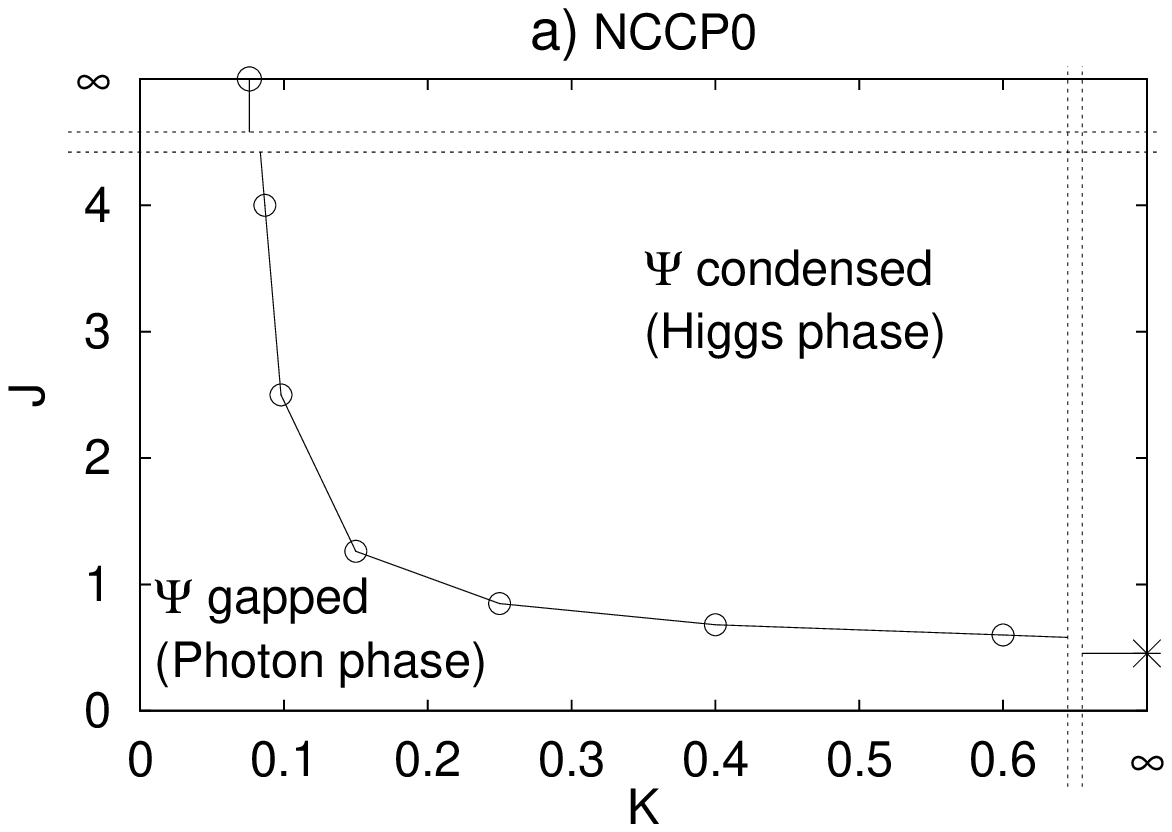}}
\vskip -2mm
\centerline{\includegraphics[width=\columnwidth]{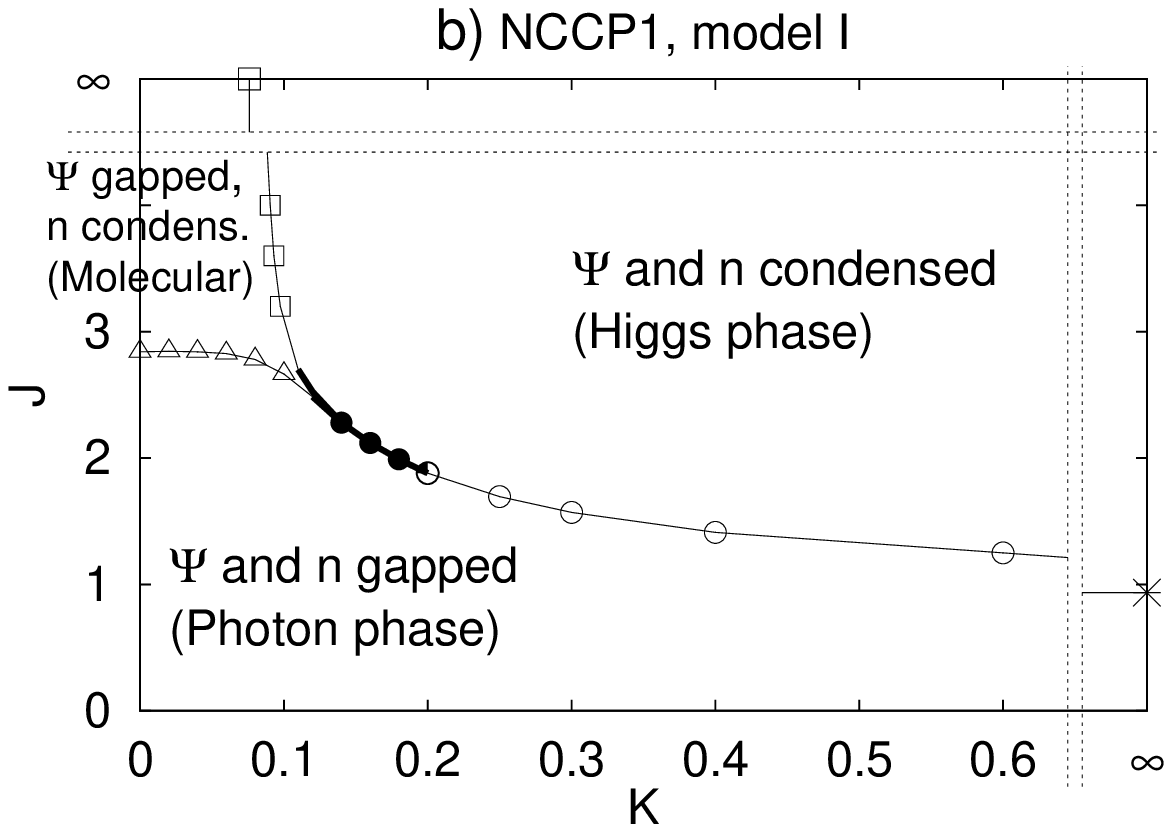}}
\vskip -2mm
\centerline{\includegraphics[width=\columnwidth]{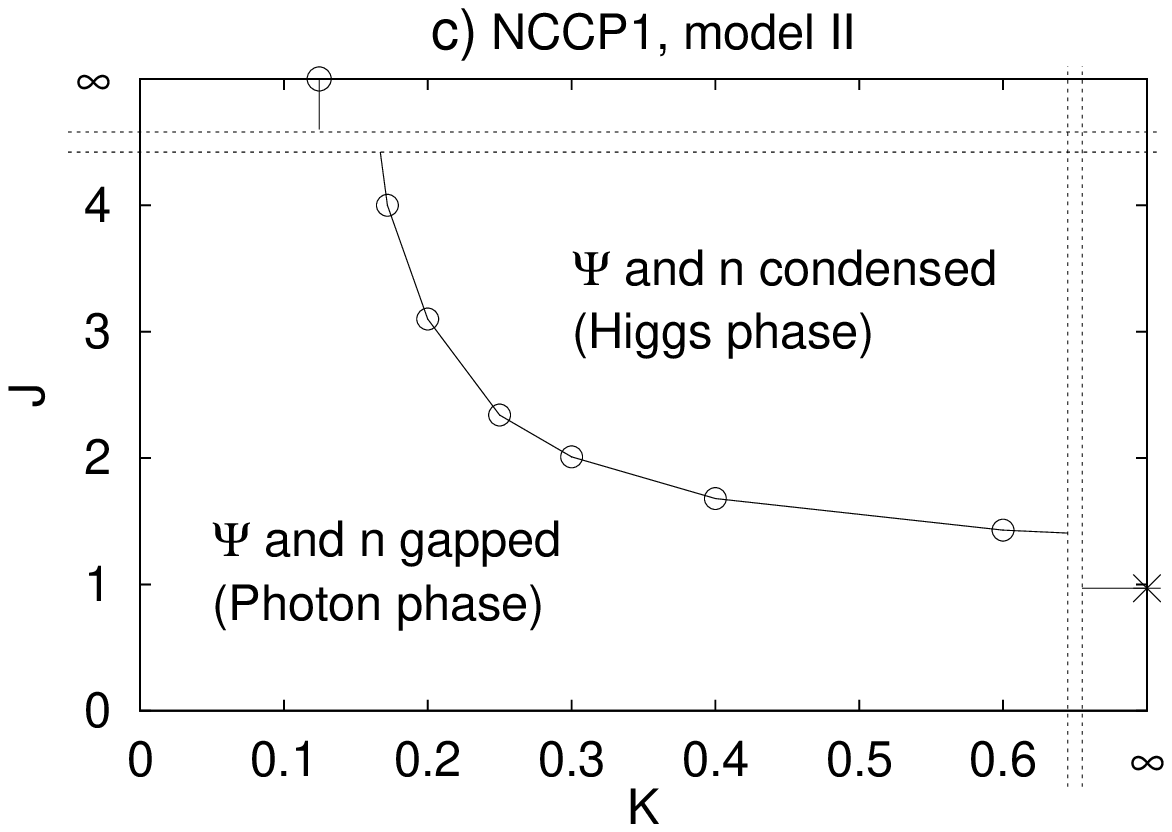}}
\vskip -2mm
\caption{Phase diagrams of the three models.
In the NCCP$^0$ model, $\Psi = e^{i\theta}$, while in the NCCP$^1$ case,
${\bm \Psi} = {\bm z} = (z_\up, z_\dn)$ is a two-component field.
a) The transition between the Photon and Higgs phases is inverted XY
and is denoted with open symbols, except at ${\cal K}=\infty$
where it is direct XY (denoted with star).
b) NCCP$^1$ model I has a rich phase diagram which contains
also the Molecular phase where ${\bm \Psi}$ is gapped but $\n$ orders.
Our main interest is on the Photon to Higgs phase transition;
the points where it is continuous are denoted with open circles,
while in the vicinity where the three phases meet the transition
becomes first order and is denoted with filled circles.
c) The Molecular phase is eliminated in the NCCP$^1$ model II,
allowing to focus solely on the Photon - Higgs transition.
We find that everywhere along the phase boundary the transition is
continuous (indicated with open symbols).
}
\label{fig:phased}
\end{figure}

Figure~\ref{fig:phased} shows the phase diagrams of the three models
in the $\K - \J$ parameter space.  In each case, for small $\J$,
the matter fields are gapped and the gauge field is massless described
by the Maxwell term -- we mark this as the Photon phase.
On the other hand, when $\J$ and $\K$ are sufficiently large,
the matter fields condense and gap out the photon -- this is the
Higgs phase.
In the one-component case, there remain no gapless excitations.
In the two-component system, we can construct a gauge-invariant O(3)
vector field
\begin{equation}
\n = {\bm z}^\dagger \vec{\sigma} {\bm z} ~.
\label{ndef}
\end{equation}
This obtains an expectation value in the Higgs phase, and there appear
Goldstone modes associated with the global symmetry breaking.
The focus in this paper is on the Photon to Higgs transition driven by
the condensation of the charged matter fields.
We will also see how to view this transition coming from the
Higgs side as driven by proliferation of the ANO vortices with internal
structure.

Proceeding with more details on the phase diagrams, consider first the
NCCP$^0$ model.  At $\K = \infty$, there is no gauge field left and the
transition along this axis is 3D XY ordering of the $e^{i\theta}$ field
(in Fig.~\ref{fig:phased}a, this transition point is indicated with a
star symbol).  When $\K$ becomes finite and the matter-gauge coupling is
switched on, the universality of the transition changes.
Dasgupta and Halperin\cite{DH} argued that sufficiently far from the
$\K=\infty$ limit, the transition becomes inverted XY as summarized in
Sec.~\ref{sec:intro} in terms of the ANO vortices.
We are also mostly interested in moderate $\K$.
Using classical Monte Carlo, we studied several points along the phase
boundary and found continuous transitions in our model; these points are
indicated with open symbols.
(The limit of large but finite $\K$ was not answered in
Ref.~\onlinecite{DH}, but several scenarios were discussed.
One possibility is that the whole line is described by the
inverted XY universality.)

The picture of the transition becomes particularly simple in the
limit $\J = \infty$.  Here, the gauge field takes on $2\pi$-integer
values and the model reduces to that of conserved discrete-valued fluxes
$B_\mu = (\nabla \times a)_\mu$ with steric interactions,
\begin{equation}
Z_{{\rm NCCP}^0}^{\J=\infty}
= \sum_{{\bm B} = 2\pi \times {\rm int},
        {\bm \nabla} \cdot {\bm B} = 0,
        {\bm B}_{\rm tot} = 0}
\exp \left[ -\frac{\K}{2} \sum_\Box B^2 \right] ~.
\label{S_extrmnccp0}
\end{equation}
This is the familiar loop representation of the 3D XY model and
has been studied in great detail in the literature.  Note that in our
ensemble derived from the gauge model the total flux is constrained
to be zero, but this is not important in the thermodynamic limit.
There are only small loops for $\K > \K_c = 0.07606$, but they
proliferate for $\K < \K_c$.  The transition is 3D XY in terms of these
short-range interacting loops, which are precisely the ANO vortices
carrying quantized flux.  These loops are gapped on the Higgs side and
condense upon reducing $\K$ and $\J$, which is why the transition is
referred to as `inverted XY' (as opposed to `direct XY' transition
at $\K = \infty$ where currents conjugate to $e^{i\theta}$ proliferate
upon increasing $\J$).
The universality of the transition remains unchanged also away from the
$\J = \infty$ limit.  Indeed, one can see that finite $\J$ only leads to
short-range modifications of the ANO vortex interactions and therefore
does not change the nature of the transition.
To conclude the description of the NCCP$^0$ phase diagram, we readily
see that there are no transitions along the $\K=0$ or $\J=0$ axes in
Fig.~\ref{fig:phased}a.

Consider now the NCCP$^1$ model I, whose phase diagram is shown
in Fig.~\ref{fig:phased}b.
Similarly to the NCCP$^0$ case, there is no transition along the
$\J=0$ axis.  Along the $\K=\infty$ line, the gauge field is again
absent and the transition is in the O(4) universality class,
but the universality changes once $\K$ becomes finite.
The region where we find the transition is continuous in a new
universality class is marked with open symbols in the figure
(details of the numerical study of the transitions are given
in Sec.~\ref{sec:transitions}).

Unlike the one-component case, this two-component model has an additional
phase at small $\K$ and large $\J$, which, while interesting on its own,
does not bear on the universality class of the Higgs transition that we
are interested in.  For completeness and in order to better understand
the numerics in this model, let us describe the new phase, which we
mark as Molecular.  Here, the charged field ${\bm z}$ is gapped but the
charge-neutral field $\n$ obtains an expectation value,
and the O(3) order and the photon coexist.
This molecular condensate appears because of the attraction between
oppositely charged ${\bm z}^\dagger$ and ${\bm z}$ fields.
Explicitly, consider the $\K=0$ axis.  The gauge field can be
integrated out completely, with the resulting action for
the ${\bm z}$'s
\begin{equation}
S_{{\rm NCCP}^1, \text{model I}}^{\K=0} [{\bm z}] =
- \sum_{i,\mu} \ln I_0(\J |{\bm z}_i^\dagger {\bm z}_{i+\mu}|) ~.
\label{nccp1_K0}
\end{equation}
Since
\begin{equation}
|{\bm z}_i^\dagger {\bm z}_j| = \sqrt{\frac{1 + \n_i \cdot \n_j}{2}} ~,
\end{equation}
there is thus a ferromagnetic interaction between neighboring spins
that drives an O(3) ordering transition upon increasing $\J$
along the $\K=0$ axis.
The transition remains in the same universality also for small $\K$.
Since the condensate is neutral, the photon remains free and is a
harmless spectator across this Photon - Molecular phase transition.

Consider now the transition from the Molecular to the Higgs phase
at large $\J$.  This is similar to the one-component transition
described earlier.
Indeed, if we take an ordered state of the $\n$'s, the remaining
U(1) phase degrees of freedom of ${\bm z}$'s are coupled to the gauge
field in the same manner as the $e^{i\theta}$'s in Eq.~(\ref{nccp0}),
and the Molecular to Higgs phase transition is the condensation of this
charged field.  In particular, the transition at $\J = \infty$ occurs
at the same $\K_c$ and is in the inverted 3D XY class; the transition
is expected to remain in this universality also away from the
$\J = \infty$ limit.
The Goldstone modes of the broken global O(3) are present both in the
Molecular and Higgs phases and are harmless spectators across this
transition.

Having discussed the three phases, consider now the region where
they meet -- the `fork' in the phase diagram.
A mean field argument\cite{Kuklov} summarized in Appendix~\ref{app:mf}
shows that in the immediate vicinity of the fork the Photon to Higgs and
Molecular to Higgs transitions become first order,
see Fig.~\ref{fig:mfphased}.
The Photon to Molecular transition, on the other hand,
remains continuous.  Focusing on the Photon to Higgs transition,
the first order line terminates at a so-called tricritical point,
beyond which the transition is continuous.
We indeed find similar picture in the Monte Carlo simulations of the
model, where the transition is second-order for the $\K > 0.2$ points
in Fig.~\ref{fig:phased}, and this is the region that we are most
interested in.
However, as we will describe in Sec.~\ref{subsec:nccp1_trends},
a detailed study of the universality is hampered by the fact that
this region is bordered by two different critical points that
are not related to the fixed point of our transition:
On one hand, the system needs to be sufficiently away from the
$\K=\infty$ limit to depart from the O(4) universality class.
On the other hand, it also needs to be far from the first-order
end-point that imposes its own crossovers before the system flows
to the fixed point that controls the transition we are interested in.
Note that the renormalization group flows that we have in mind are in a
more general space than the $\K - \J$ parameter space of the model,
so the above scenario is consistent.

Because of these complications, the NCCP$^1$ model I, while simple to
formulate, requires good understanding of all crossovers before
interpreting the results.  We do not have such a complete control
to be able to reliably extract the critical indices,
but we will show representative points on the phase boundary where
numerics is strongly in favor of the second-order character which is
``more continuous'' than in the NCCP$^0$ case.
We will further argue for the hypothesis of the tricritical point in
this model in Sec.~\ref{subsec:nccp1_trends}.

The universality of the Higgs transition that we want to study
is independent of additional short-range interactions as long
as the transition remains continuous.
In this respect, to better focus on the scaling properties,
we also consider the NCCP$^1$ model II, Eq.~(\ref{nccp1wAF}),
where we have added some antiferromagnetic interaction between
neighboring spins that hinders the ordering of the $\n$'s.
In fact, with the hindsight of Eq.~(\ref{nccp1_K0}), the chosen
interaction precisely compensates the ferromagnetic interaction
in the original model I along the $\K = 0$ axis.
In particular, the model II has no phase transition along this axis.
We actually find that the Molecular phase is eliminated in this model,
and the resulting phase diagram is displayed in Fig.~\ref{fig:phased}c.
The transition remains continuous all the way to the $\J = \infty$ limit,
and the phase diagram in the $\K - \J$ plane looks similar to the
NCCP$^0$ case.

Thus, in the NCCP$^1$ model II, we avoid the complicating crossovers
caused by the possible tricritical point in the vicinity of the
Molecular phase.
We only need to worry about being sufficiently away from the
$\K = \infty$ limit, and this is similar to the NCCP$^0$ case that we
understand analytically and can compare numerically.
We expect that adding generic antiferromagnetic interactions between 
the $\n$'s would give a similar improvement in the NCCP$^1$ model,
suppressing the Molecular phase, but we have chosen the specific form 
to avoid possibly introducing new phases in the phase diagram. 
From a Monte Carlo study of the model II combined with a separate study 
of the $\J = \infty$ limit and an analysis of the stability away from 
this limit, we conclude that no other phase appears in this model.

Let us describe in more detail the $\J = \infty$ limit,
which we formulate as
\begin{eqnarray}
Z_{{\rm NCCP}^1, \text{model II}}^{\J=\infty}
&& = \sum_B^\prime \int D\z \, \delta(|\z|^2 = 1) \nonumber \\
&& \exp\left[-\frac{\K}{2} \sum_\Box
              \left(B + \nabla \times {\A}[\z] \right)^2
      \right] ~.
\label{S_extrmnccp1}
\end{eqnarray}
Here the primed sum is over $2\pi$-integer-valued conserved $B$-fluxes
as in Eq.~(\ref{S_extrmnccp0}).  The ensemble also contains CP$^1$
fields $\z$ residing on the sites and specifying uniquely link
variables $\A$ as follows:
\begin{eqnarray}
\label{Adef}
e^{i\A_{ij}} &=& \frac{\z_i^\dagger \z_j}{|\z_i^\dagger \z_j|} ~.
\end{eqnarray}
Note that any $2\pi$ ambiguity in the $\A$'s can be absorbed
by the $B$'s.
Crudely, we obtain Eq.~(\ref{S_extrmnccp1}) by focusing on the phases of
the link variables $\z_i^\dagger \z_j$ while completely suppressing the
role of their amplitudes, and this is precisely achieved when we
take the $\J \to \infty$ limit in the model II.

Comparing with the $\J=\infty$ limit in the NCCP$^0$ case,
the NCCP$^1$ variant can be also viewed as a loop model, but the
loops have non-trivial internal structure derived from the coupled
degrees of freedom $\z_i$.  Numerically we find that this system has a
continuous transition at $\K_c \approx 0.1245$,
see Sec.~\ref{subsec:Jinfty}, and the universality of the transition
is different from the loops with no internal structure.
We can also carry through an analysis in the model II when $\J$ is
large but finite and find that this leads only to short-range
modifications that vanish in the $\J \to \infty$ limit
(this is similar to how one analyzes the NCCP$^0$ case at large $\J$).
We then propose that the whole phase boundary in Fig.~\ref{fig:phased}c
is in the same universality as the transition found in Monte Carlo at
$\J=\infty$.

\section{Comparative study of the transitions}
\label{sec:transitions}

\subsection{Monte Carlo measurements.  Scaling analysis}
\label{subsec:method}

We first summarize our numerical measurements.
We perform classical Monte Carlo simulations using local updates
of the matter and gauge fields, where the latter are treated as
unconstrained real variables.
(We can write an explicit finitely defined statistical mechanics model
that is properly sampled by such simulations, but do not belabor
this because such details are not used here.)
In the $\J=\infty$ models, we also use geometrical worm
updates\cite{Prokofiev, Alet} of the discrete-valued conserved
fluxes $B$.
Where possible, we use multiple histogram method to interpolate
between the data points;\cite{MCBook}
the error bars are estimated by using blocking method and
by running several independent samples.

As we change the parameters in the system, we monitor, first of all,
thermal properties such as the specific heat per site $C$.
This is defined through the variance of the action,
\begin{equation}
C = \frac{\la (S - \la S \ra)^2 \ra}{L^d} ~,
\label{C}
\end{equation}
and can alert about phase transformations in the system even when
we do not know the underlying physics.
At a second-order phase transition, $C$ behaves as
$C \sim |t - t_c|^{-\alpha} = |t - t_c|^{-(2-\nu d)}$, where
in the last equation we used the relation between the specific
heat exponent $\alpha$ and the correlation length exponent $\nu$.
The singularity is rounded off in a finite system, but manifests
itself as a peak in the specific heat that evolves characteristically
with the system size.  In a finite volume $L^d$, the peak location
approaches the bulk critical point as
$t_{\rm peak} - t_c \sim L^{-1/\nu}$,
while the peak height behaves as
\begin{equation}
C_{\rm max} \sim C_0 + A L^{2/\nu - d}, \quad (\text{second order}) ~.
\label{Cmax_secondorder}
\end{equation}
For comparison, at a first order transition, the peak height
grows with the system size as
\begin{equation}
C_{\rm max} \sim L^d, \quad (\text{first order}),
\label{Cmax_firstorder}
\end{equation}
which would be measured as $\nu_{\rm eff} = 1/d = 1/3$ in a naive
analysis in 3D;
the peak location converges to the bulk critical point as
$t_{\rm peak} - t_c \sim L^{-d}$.

Thus, the behavior of the specific heat gives first indication
about the nature of the transition.  In the NCCP$^0$ model, we
anticipate a continuous transition with $\nu \simeq 0.67$ -- the
exponent $\alpha$ is then small and negative, and the singularity in
$C$ is only a cusp.
In this case, $C$ is less useful for scaling, but a good estimate
of $\nu$ can be obtained by studying the third cumulant of the action,
\cite{C3method}
\begin{equation}
C_3 = \frac{\la (S - \la S \ra)^3 \ra}{L^d} ~,
\label{C3}
\end{equation}
which can be viewed as a derivative of $C$ with respect to a
temperature-like parameter.
Near the critical point, the characteristic $C_3$
(such as the maximum value or the maximum - minimum difference)
scales as $L^{3/\nu - d}$, which is more readily detected.
This is useful in the NCCP$^1$ model as well, where, previewing
the Monte Carlo results, the thermal signature of the transition is less
singular and implies a larger exponent $\nu$.

Let us now consider direct detection of the Higgs transition.
In the Higgs phase, the gauge field is massive and the fluxes do not
fluctuate on large scales.  On the other hand, in the Photon phase,
the fluxes proliferate.  This can be measured by what we call the
dual stiffness
\begin{equation}
\rho_{\rm dual}^{\mu\mu}(q) =
\left\la \frac{|\sum b_\mu(R) e^{i q \cdot R}|^2}{(2\pi)^2 L^3}
\right\ra ~.
\label{rhodual}
\end{equation}
Here the summation is over all plackets oriented perpendicular to
$\hat \mu$, $b_\mu = (\nabla \times a)_\mu$ is the corresponding flux,
and $R$ is the coordinate of the placket.  We consider the stiffness
at a wavevector $q$ and are interested in the small $q$ limit.
In an infinite system, $\rhodual$ is zero in the Higgs phase and is
finite in the Photon phase.  If the fluxes are viewed as conserved
currents, $\rhodual$ is essentially the superfluid stiffness of this
current loop system.

Recall also that in the NCCP$^0$ case at $\J = \infty$, the fluxes
$b/(2\pi)$ become integer-valued (the corresponding loops are
the ANO vortex lines).
$\rhodual$ is then precisely the conventional superfluid stiffness
for this loop system (which also explains our normalization in
Eq.~\ref{rhodual}).  The loop model is usually studied with no
restriction on the total flux, and one usually measures the stiffness
at $q=0$.  This is not available in our setup: since we are working with
the gauge fields and periodic boundary conditions, the total flux is
constrained to be zero.  However, we can measure the stiffness at a
nonzero $q \sim 1/L$, and in the present work we use
$\rho_{\rm dual}^{xx}$ and $\rho_{\rm dual}^{yy}$ at the smallest
$q_{\rm min}=(0, 0, 2\pi/L)$.
In a finite system, this observable is less sharp than $\rho_{q=0}$.
Thus, in the phase with only small loops, there is a contribution to
$\rho_q$ of order $q^2 \sim 1/L^2$, while the $q=0$ stiffness would
vanish exponentially with $L$.  Still, we find that this observable
works well.  The NCCP$^0$ model allows us to develop such
experience and test our protocols before studying the NCCP$^1$ case.

At a second-order transition, we expect the product
$\rhodualqmin \cdot L$ to be universal, while near criticality
\begin{equation}
\rhodual (q_{\rm min}; t, L) \cdot L= r(\delta L^{1/\nu}) ~,
\label{rhod_scaling}
\end{equation}
where $\delta = t-t_c$ measures deviation from the critical point and
$r(x)$ is a scaling function.  We plot the $\rhodual \cdot L$ curves
and look at their crossings and thus estimate the location of the
critical point.  We also extract the correlation length exponent $\nu$
by applying the above scaling form.  We try different approaches.
One is to consider a derivative with respect to a temperature-like
parameter in the system (such derivatives can be calculated accurately
in the Monte Carlo process).  At the critical point we expect
\begin{equation}
\frac{d(\rhodual \cdot L)}{d t}|_{\rm crit} \sim L^{1/\nu} ~,
\label{deriv_scaling}
\end{equation}
which can be used to estimate $\nu$.  This approach is sensitive to
our uncertainty in the location of the critical point and is not
very robust: e.g., the critical such curve does not separate the ones
on the two sides of the transition but eventually should go above
all of them.
A different procedure is to consider the maximal value of the derivative
for each $L$.  Yet another approach is to scale the full
$\rhodual \cdot L$ curves near criticality.
We are using all such approaches, but a crude look at the derivatives
is simple and will already be sufficient for our comparison of the
NCCP$^0$ and NCCP$^1$ models -- our main message will be that $\nu$ is
larger in the latter case.

The discussed thermal measures and the dual stiffness work in any
matter-gauge system.  In the NCCP$^1$ case, we also have the field $\n$,
Eq.~(\ref{ndef}), which is a local physical observable in the matter
sector.  We measure the magnetization associated with the O(3) ordering,
\begin{equation}
\vec{m} = \frac{1}{L^d} \sum_i \vec{n}_i,
\label{m}
\end{equation}
and estimate the critical point from the crossings of the corresponding
Binder cumulant ratio
\begin{equation}
g = \frac{\la |\vec{M}|^4 \ra}{\la |\vec{M}|^2 \ra^2} ~.
\label{g}
\end{equation}
Similarly to Eq.~(\ref{rhod_scaling}), we expect the following scaling
form near the critical point:
\begin{equation}
g (t, L) = g(\delta L^{1/\nu}) ~,
\label{g_scaling}
\end{equation}
which we analyze in the same spirit as the $\rhodual \cdot L$.

We characterize the O(3) ordering by the magnetization exponent
$\beta$: $m \sim \delta^{\beta}$.  This gives the following
finite-size scaling ansatz
\begin{equation}
m(t, L) = L^{-\beta/\nu} f(\delta L^{1/\nu}) ~.
\label{m_scaling}
\end{equation}
In particular, right at the critical point we expect
\begin{equation}
m|_{\rm crit} \sim L^{-\beta/\nu} = L^{-(1+\eta)/2},
\label{mcrit_scaling}
\end{equation}
where the last equation defines exponent $\eta$ in 3D.
By plotting $m$ vs $L$ on a log-log plot, the critical curve is a
straight line that separates the curves on the ordered and
disordered sides.
This analysis independently checks the location of the critical point
and gives an estimate of $\beta/\nu$.

Finally, we also observe the transition by monitoring the
helicity modulus $\Upsilon$ for twisting the O(3) vector field $\n$.
This can be calculated in the standard way during the same Monte Carlo
process, and we give detailed expressions for the two NCCP$^1$ models in
Appendix~\ref{app:helicity}.
The helicity modulus is also a measure of the ordering in the O(3)
matter sector, but has a somewhat different character from the
magnetization and is more like a superfluid stiffness in a boson system.
The scaling of the helicity modulus at a second-order transition is
similar to the dual stiffness described earlier.
In particular, we expect universal crossings of the $\Upsilon \cdot L$
curves, which is yet another way of locating the transition in our
finite-size study.
These measurements can also be compared with other studies of
such matter-gauge systems,\cite{Kuklov} and also with studies of
quantum Hamiltonians whose critical behavior may be described by
such field theories.\cite{XYring, SU2ring1, SU2ring2}

\subsection{Representative study of the transitions in the
three models at K=0.4}
\label{subsec:K40}

Here we present Monte Carlo results at $\K = 0.4$ in the three models.
This is sufficiently away from the $\K = \infty$ limit -- for example,
the critical coupling $\J_c(\K=0.4)$ is at least 50\% larger than
$\J_c(\K=\infty)$ in each case:
$\J_c(0.4) = 0.682$ vs $\J_c(\infty) = 0.4541$ in the NCCP$^0$ model;
$\J_c(0.4) = 1.412$ vs $\J_c(\infty) = 0.9358$ in the NCCP$^1$ model I;
and $\J_c(0.4) = 1.68$ vs $\J_c(\infty) = 0.98$ in the NCCP$^1$ model II.
Furthermore, the transition is clearly observed in the behavior
of the gauge field as detected by the dual stiffness --
we see a convergence of the crossings of $\rho_{\rm dual} \cdot L$
for our system sizes.
In the NCCP$^1$ models, we also see convergence of the Binder ratio
crossings and the $\Upsilon \cdot L$ crossings, which detect the transition in the matter sector.
Recall that the gauge sector is completely absent in the $\K=\infty$
limit, and indeed at large $\K$ the ``critical points'' defined by the
``gauge'' and ``matter'' measures start far apart for our system sizes,
but are already close at $\K=0.4$.  We expect comparable finite size
effects in the three systems as far as the development of the true Higgs
criticality is concerned.

We will discuss the observed trends along the Photon - Higgs
phase boundary in Secs.~\ref{subsec:nccp0_nccp1wAF_trends},
\ref{subsec:nccp1_trends}.
The representative point $\K=0.4$ was chosen to be sufficiently away
from the first-order region in the NCCP$^1$ model I, with the hope
that it is not affected strongly by crossovers near the tricritical
point.  We argue here that the transition is continuous in all three
models at $\K=0.4$.  In the NCCP$^0$ case, this is expected
theoretically and is supported numerically.  In all measured respects,
the transition in either of the NCCP$^1$ models at $\K=0.4$ looks
more continuous than NCCP$^0$.

\begin{figure}
\centerline{\includegraphics[width=\columnwidth]{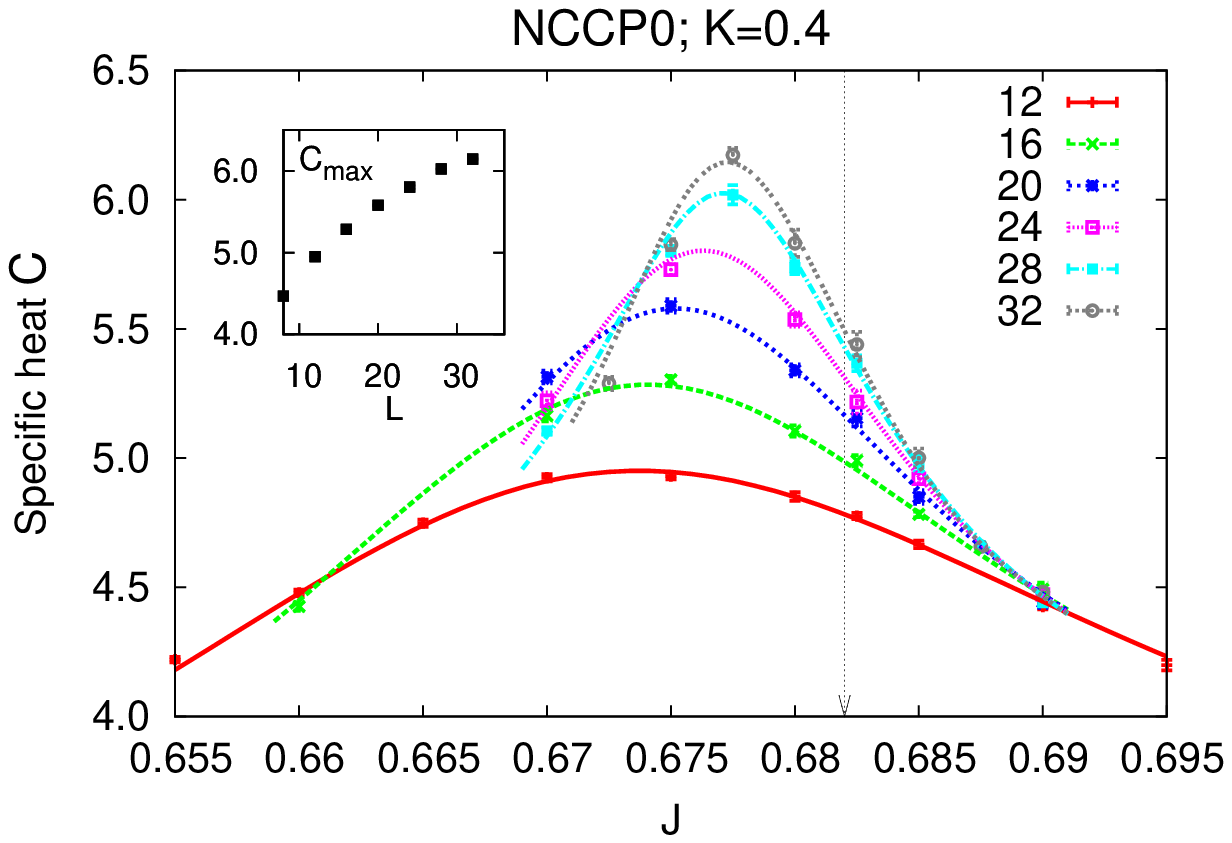}}
\centerline{\includegraphics[width=\columnwidth]{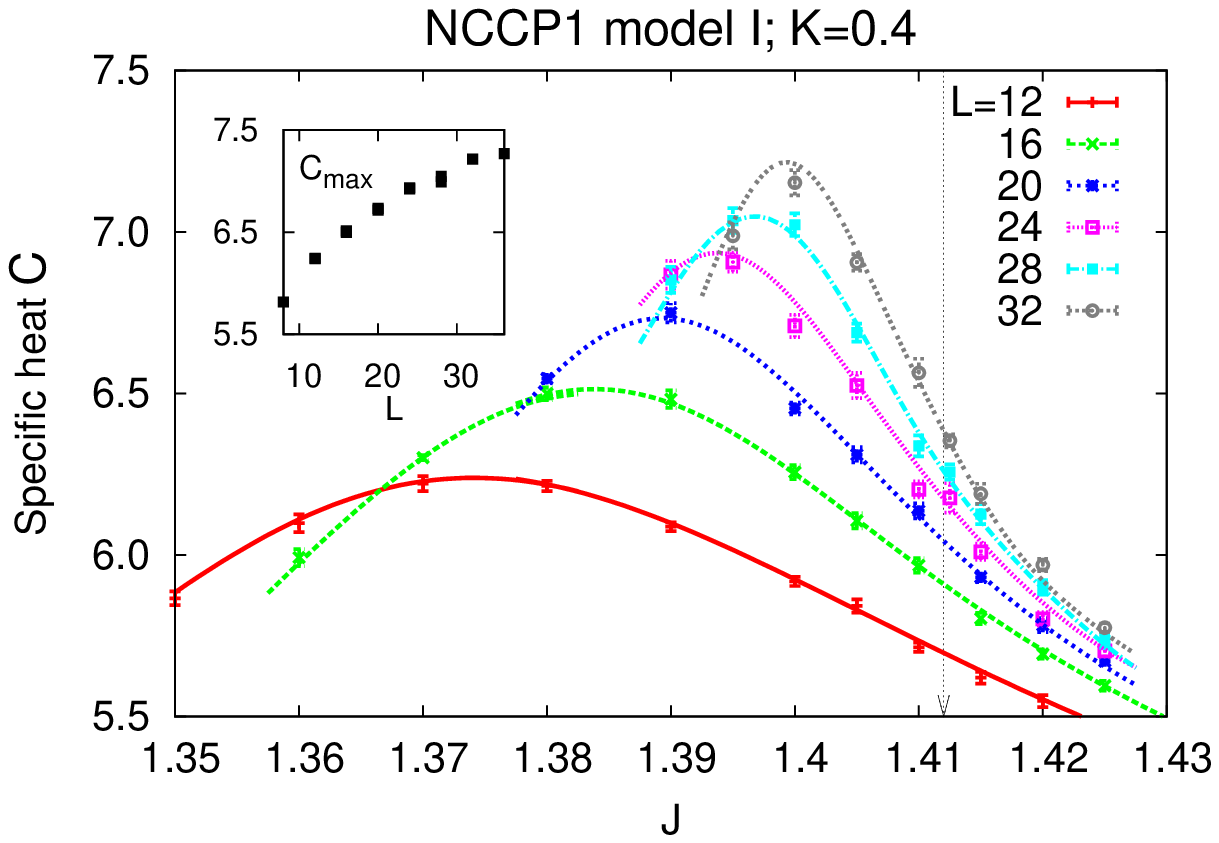}}
\centerline{\includegraphics[width=\columnwidth]{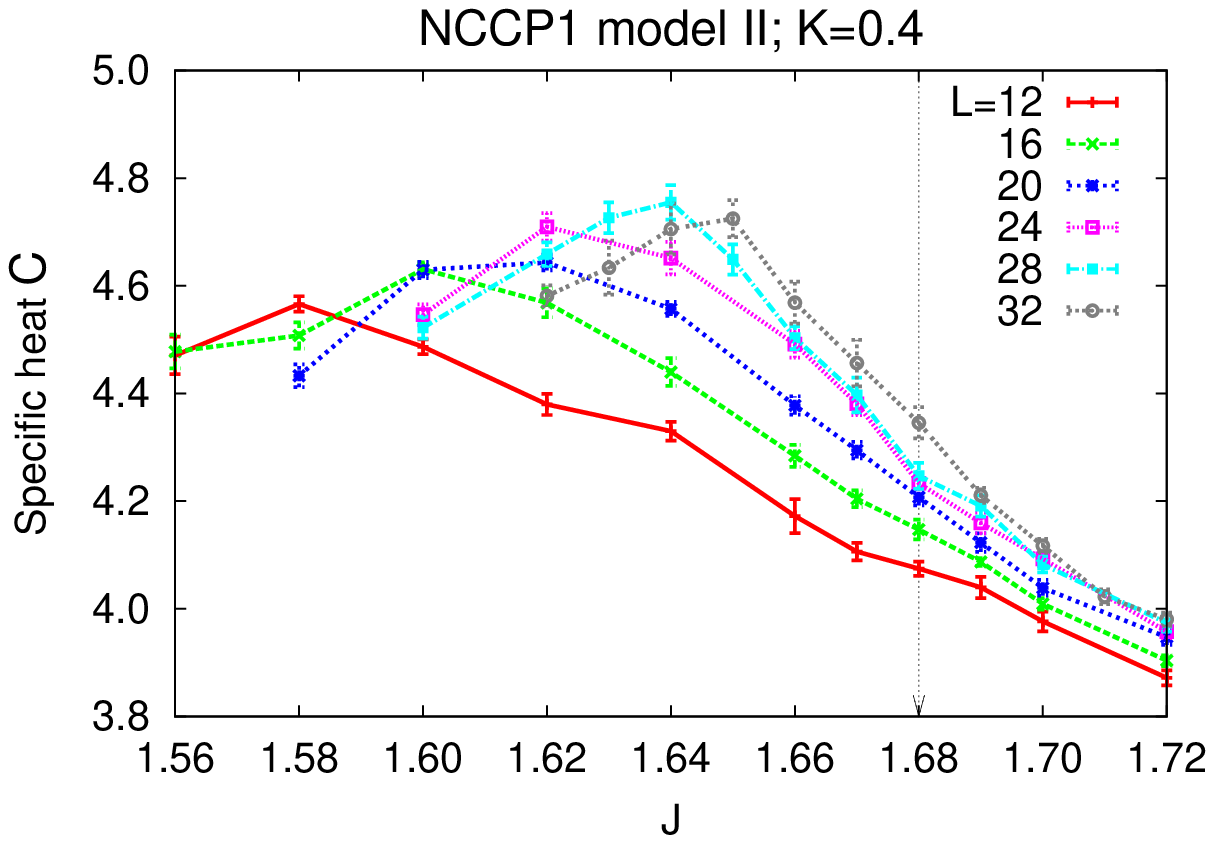}}
\caption{(color online).
Specific heat $C$ in the three models at ${\cal K}=0.4$
varying ${\cal J}$ near the transition.  System sizes are $L = 12 - 32$.
In the top and middle panels, the lines are obtained by multiple
histogram method, which provides a systematic interpolation between
data points; the corresponding peak values $C_{\rm max}$ are
plotted in the insets and show sublinear growth with $L$.
In the bottom panel, the lines are only guides to the eye connecting
nearby data points; in this case, the peak value changes too little
to be extracted reliably.
In each panel, the vertical arrow shows the bulk critical point
${\cal J}_c$ estimated from Figs.~\ref{fig:rhodualQ001xL_K40}
and \ref{fig:BM_K40}.
The specific heat peak moves towards this point gradually and the
developing singularity is weak, which is a direct evidence of the
second-order nature of the transition.
}
\label{fig:C_K40}
\end{figure}

Figure~\ref{fig:C_K40} shows the specific heat per site in the
three models.  In each case, we see a peak whose position moves towards
the estimated critical point and whose height increases gradually.
The slow evolution of the position and the weak (at most sublinear)
growth of $C_{\rm max}$ with the system size are indicative of the
continuous nature of the transition and rather large correlation
length exponent $\nu$ in each case.
Fitting $C_{\rm max}$ to Eq.~(\ref{Cmax_secondorder}), we estimate
$\nu \approx 0.65 - 0.7$ in the NCCP$^0$ case, which is close to the
accepted value in the (inverted) 3D XY.
We find a somewhat larger $\nu \approx 0.70$ in the
NCCP$^1$ model I case.
In the NCCP$^1$ model II, the growth of the peak height with the system
size is very weak and we do not attempt to analyze $C_{\rm max}$.
We also measure the third cumulant $C_3$, Eq.~(\ref{C3}), and
analyzing this data gives estimates of $\nu$ consistent with the above.
In the three systems at $\K=0.4$, we do not see any tendency towards
first order behavior such as Eq.~(\ref{Cmax_firstorder}).
Furthermore, energy histograms for system sizes up to $L = 36$
show one mode and no sign of any structure developing.
From all such measures, the transition in the presented NCCP$^1$ systems
has weaker thermal singularities than in the NCCP$^0$ case.

\begin{figure}
\centerline{\includegraphics[width=\columnwidth]{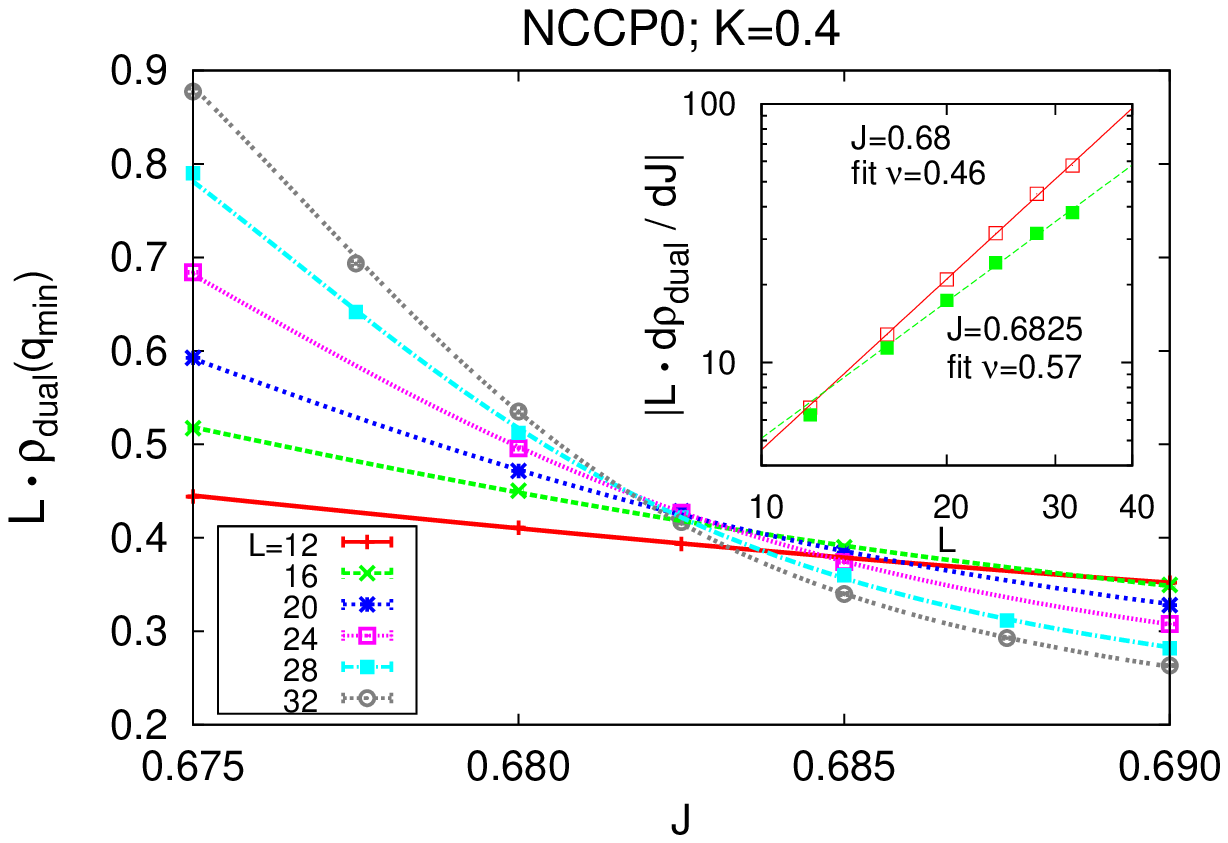}}
\centerline{\includegraphics[width=\columnwidth]{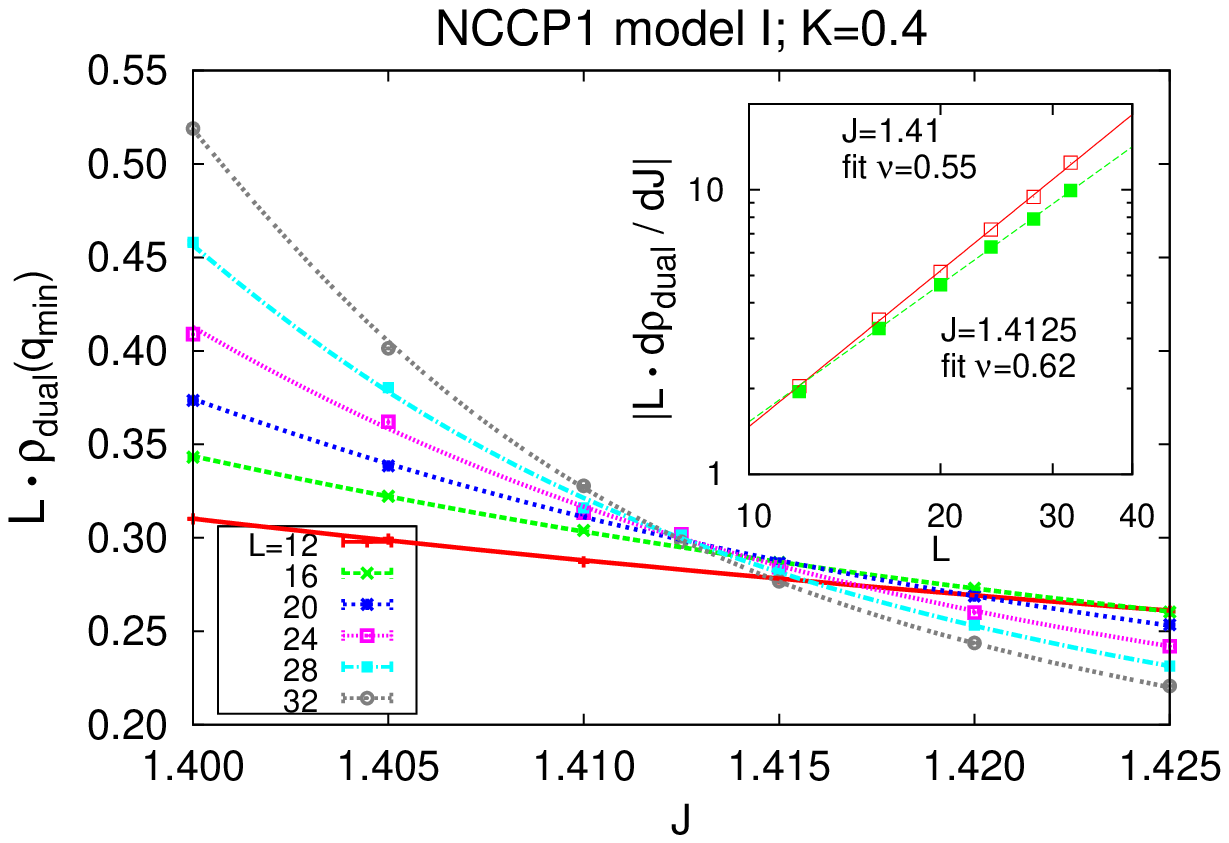}}
\centerline{\includegraphics[width=\columnwidth]{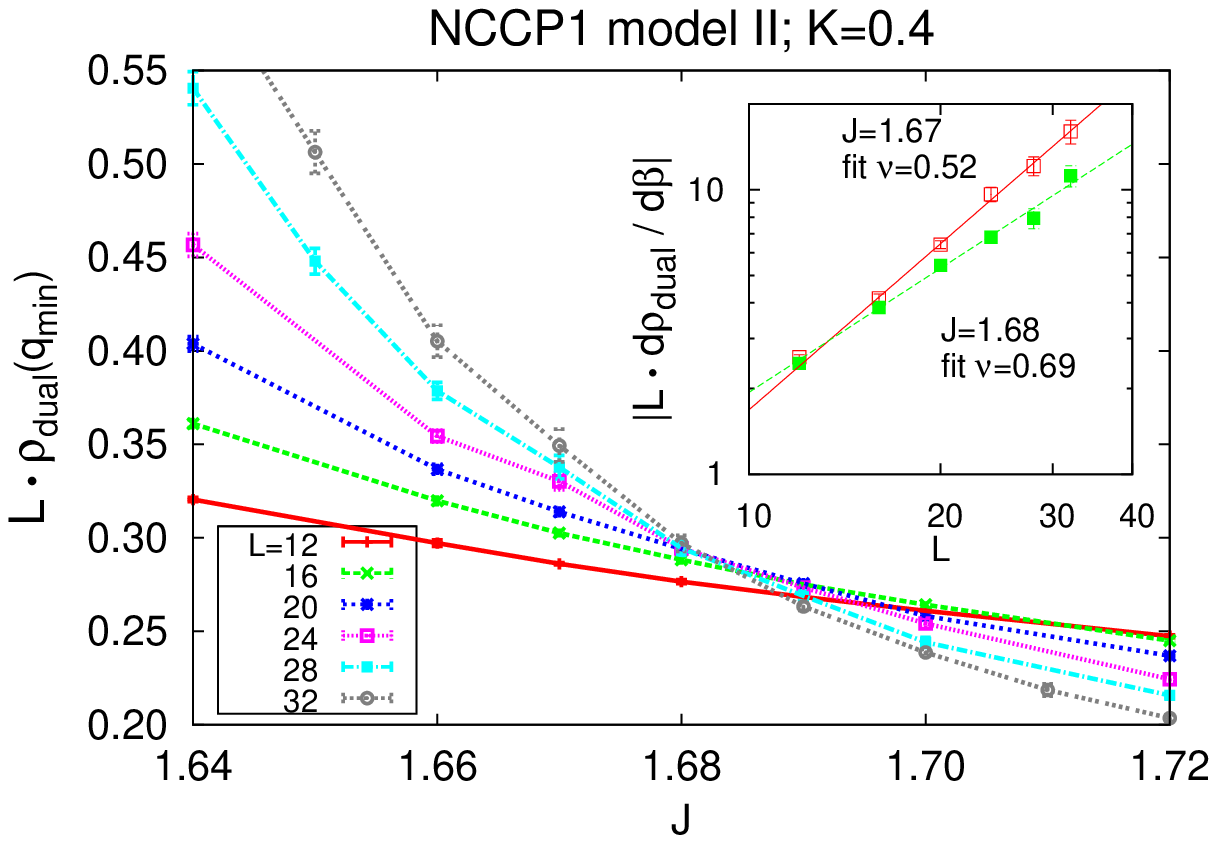}}
\caption{(color online).
$\rhodualqmin \cdot L$ in the three models at $\K = 0.4$.
The crossings converge quickly towards
$\J_c = 0.682$ in the NCCP$^0$ model (top panel),
$\J_c = 1.412$ in the NCCP$^1$ model I (middle panel),
and $\J_c = 1.68$ in the NCCP$^1$ model II (bottom panel).
In the last two cases, we can check the convergence against the
Binder ratio crossings, Fig.~\ref{fig:BM_K40}, which move in the
opposite direction.  In each panel, the inset summarizes naive estimates
of the correlation length exponent $\nu$ applying
Eq.~(\ref{deriv_scaling}) to the derivative data at points bordering
the transition -- see text for details.
}
\label{fig:rhodualQ001xL_K40}
\end{figure}

Let us describe how we locate the transitions more accurately.
Figure~\ref{fig:rhodualQ001xL_K40} shows the crossings of
$\rhodualqmin \cdot L$, which approach the critical point from the
opposite direction compared with the $C_{\rm max}$ positions.
The crossings converge quickly for our system sizes and thus
allow an accurate determination of $\J_c$.

\begin{figure}
\centerline{\includegraphics[width=\columnwidth]{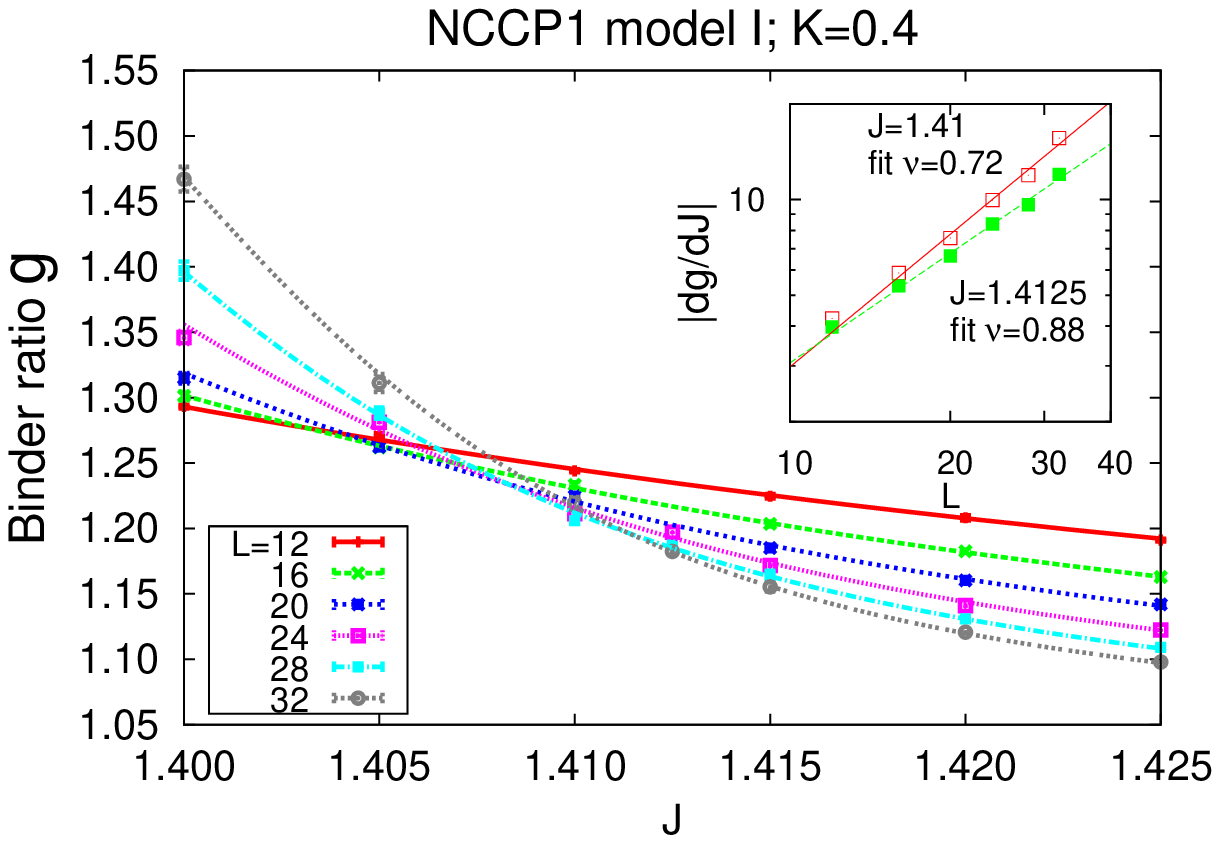}}
\centerline{\includegraphics[width=\columnwidth]{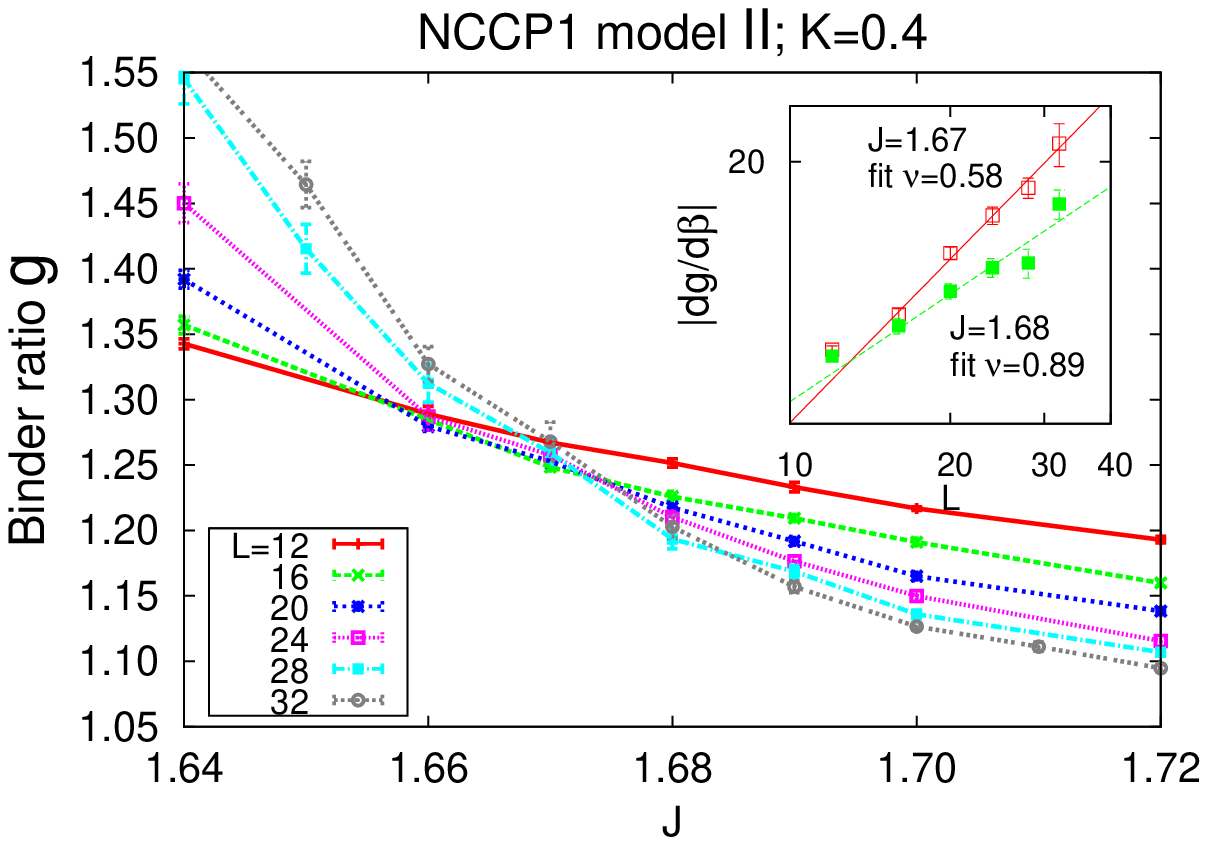}}
\vskip -2mm
\caption{(color online).
Binder ratios in the NCCP$^1$ models I and II at $\K = 0.4$.
The crossings bound the transition point from the opposite side compared
with the $\rhodualqmin \cdot L$ crossings,
Fig.~\ref{fig:rhodualQ001xL_K40}.
In the insets, we show naive estimates of $\nu$ fitting the derivative
of the Binder ratio to the form $\sim L^{1/\nu}$ at points bordering the
transition.
}
\label{fig:BM_K40}
\end{figure}

In the NCCP$^1$ systems, we also observe the transition directly in the
matter sector by measuring the magnetization $\vec{m}$, Eq.~(\ref{m}).
The corresponding Binder ratios are shown in Fig.~\ref{fig:BM_K40}
for the models I and II.  Comparing with the appropriate panels in
Fig.~\ref{fig:rhodualQ001xL_K40}, we see that the Binder crossings
move in the opposite direction to the $\rhodualqmin \cdot L$ crossings,
and the two measures effectively bound the location of the transition
to a narrow region.

\begin{figure}
\centerline{\includegraphics[width=\columnwidth]{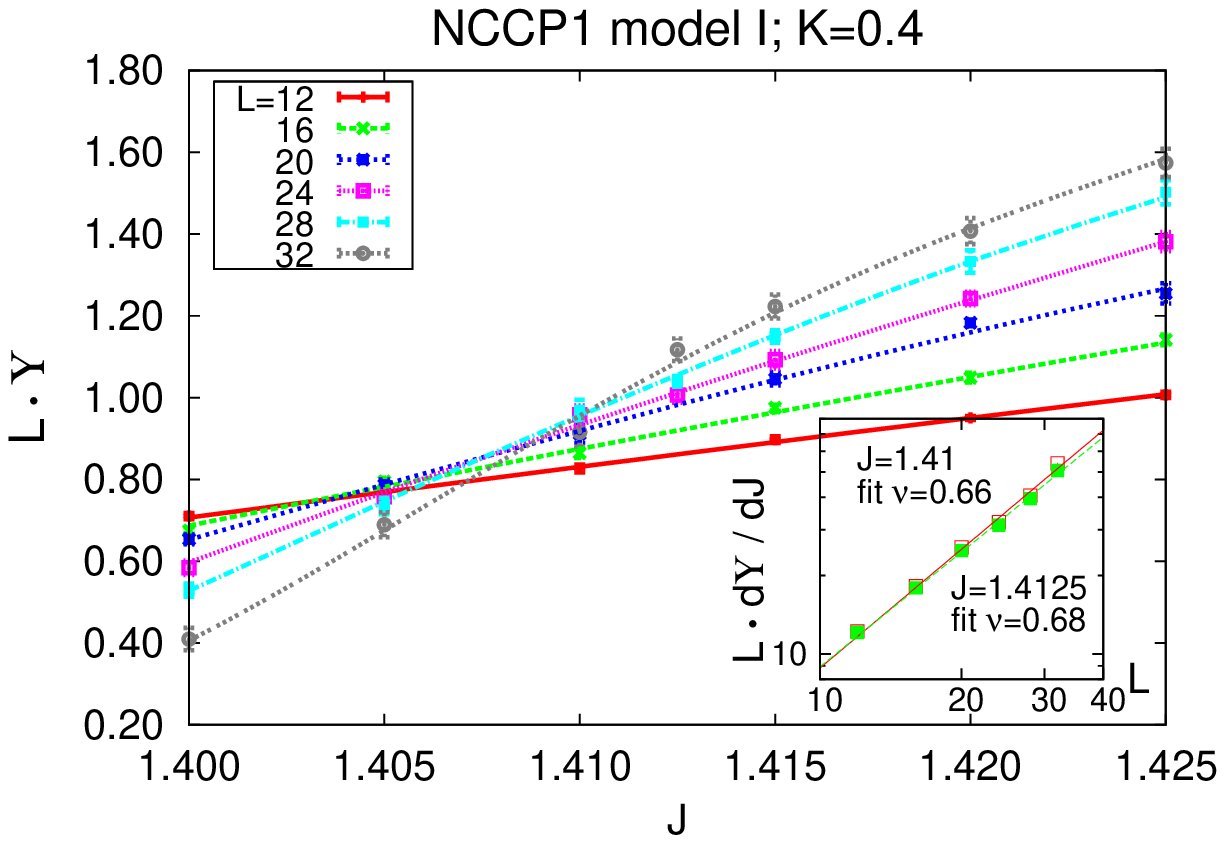}}
\centerline{\includegraphics[width=\columnwidth]{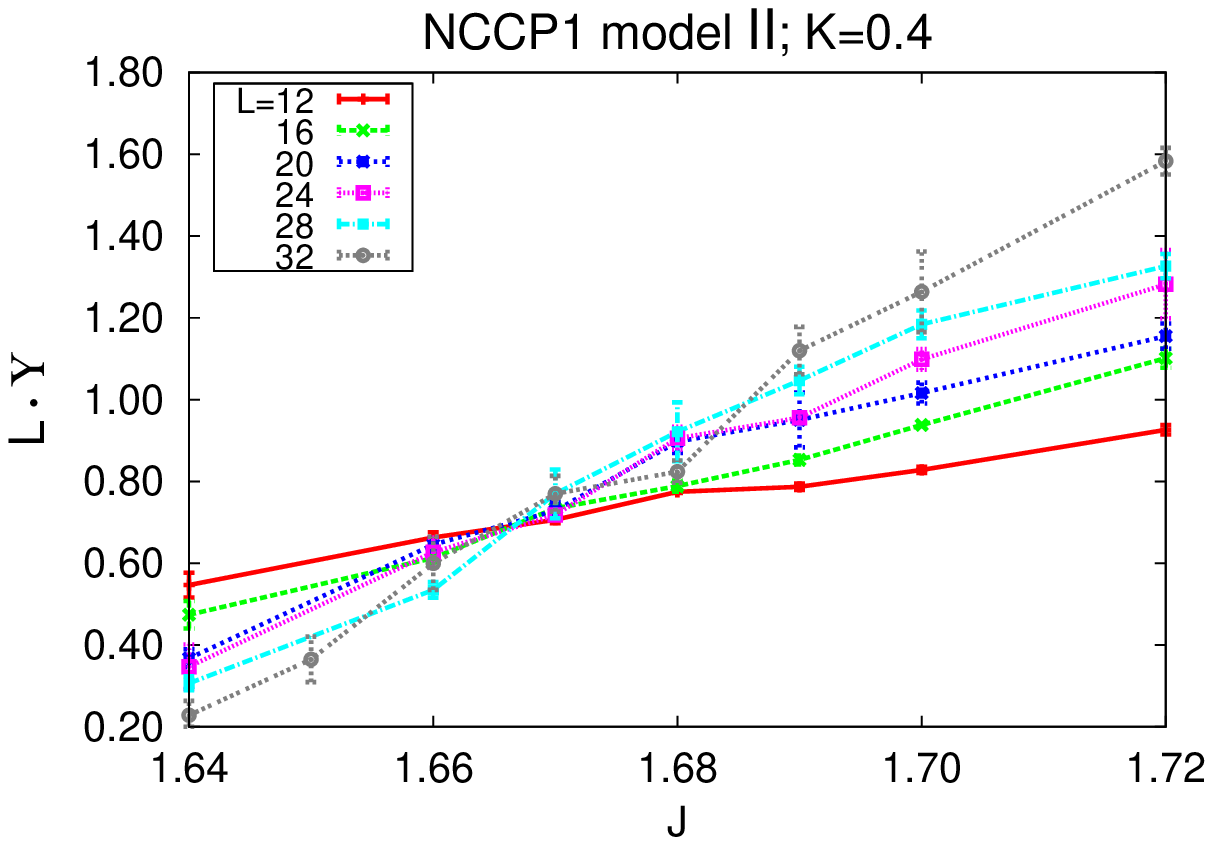}}
\vskip -2mm
\caption{(color online).
$\Upsilon \cdot L$ in the NCCP$^1$ models I and II at $\K = 0.4$.
The crossings move in the same direction as the Binder ratio crossings,
Fig.~\ref{fig:BM_K40}, and bound the transition point from the
opposite side compared with the $\rhodualqmin \cdot L$ crossings,
Fig.~\ref{fig:rhodualQ001xL_K40}.  The inset in the model I case shows
scaling analysis of the derivative similar to the analysis in
Fig.~\ref{fig:rhodualQ001xL_K40} (no such derivative data was
measured in the model II).
}
\label{fig:rhosL_K40}
\end{figure}

Finally, Fig.~\ref{fig:rhosL_K40} shows the crossings of
$\Upsilon \cdot L$ in the two NCCP$^1$ models, providing an alternative
detection of the transition in the matter sector.
The crossings converge towards the bulk critical point similarly to
the Binder crossings and also help to bound the transition region.

Having estimated the critical points, let us now discuss scaling in
the three models.  In the panels in Figs.~\ref{fig:rhodualQ001xL_K40},
\ref{fig:BM_K40}, and \ref{fig:rhosL_K40}, we also show the
corresponding derivatives evaluated at points bounding the
transition\cite{derivs_footnote} vs the system size $L$.
In each case, we perform simple linear fits corresponding to
Eq.~(\ref{deriv_scaling}) on the log-log plot and list the extracted
values of $\nu$.
This procedure is not very reliable and does not necessarily bound the
exponent, but it gives a crude picture of the results we obtain doing
more elaborate scalings.

In the NCCP$^0$ case, such scalings of $\rhodualqmin \cdot L$ suggest
$\nu$ between $0.5$ and $0.55$, which we interpret as some effective
value on our length scales.  This is significantly different from the
expected $\nu \simeq 0.67$, but when we look closer, we do see the
corresponding plots curving in the direction of such larger $\nu$.
Thus, naive scaling of $\rhodualqmin \cdot L$ in the one-component
matter-gauge system at this $\K$ suffers from strong finite size effects.
We could in principle improve our estimates of $\nu$ by performing more
complex fits including corrections to scaling, but in the absence of
good control over these, we do not venture in this direction.
Instead, in the spirit of the comparative study, we should be prepared
for similar effects also in the two-component system.

In the NCCP$^1$ model I, such scalings of $\rhodualqmin \cdot L$
show effective $\nu$ between $0.55$ and $0.65$.
This is larger than the effective $\nu$ in the above NCCP$^0$ case,
suggesting that the true $\nu$ in the two-component system is larger
than in the one-component system.
From the scaling of the Binder ratio, of the $\Upsilon \cdot L$, and
of other observables in the matter sector, we estimate $\nu$ between
$0.65$ and $0.7$ (somewhat smaller than the crude values in the
top panel in Fig.~\ref{fig:BM_K40}), which further supports our proposal
$\nu_{{\rm NCCP}^1} > \nu_{{\rm NCCP}^0}$.
We warn, however, that our scalings in the model I at $\K=0.4$
may already be affected by the proximity of the Molecular phase
-- see the discussion of the trends along the Photon - Higgs phase
boundary in Sec.~\ref{subsec:nccp1_trends}.

Turning to the NCCP$^1$ model II, the scalings of $\rhodualqmin \cdot L$
consistently suggest $\nu$ around $0.7$ (somewhat larger than the crude
estimates in the bottom panel in Fig.~\ref{fig:rhodualQ001xL_K40}),
while the scalings of the Binder ratio suggest $\nu$ between
$0.7$ and $0.75$.  This system does not have any Molecular phase
nearby and, being sufficiently away from the $\K=\infty$ limit,
is closer to the true two-component Higgs criticality than the model I.
The $\Upsilon \cdot L$ data in the model II is rather noisy, as one
see from the corresponding panel in Fig.~\ref{fig:rhosL_K40}.
This model is harder to simulate than model I: e.g., we cannot use
local heat bath updates of the matter fields and we cannot use
multi-histogram reweighting; there is also significant complexity
in the expression for $\Upsilon$, Eq.~(\ref{Upsilon}).
Still, we see that $\Upsilon \cdot L$ shows reasonable crossings
and is bounded by value of order $1$ at criticality with no
significant drifts.  Here, it is helpful to have the bounds on
the critical $\J_c$ from the more accurate $\rhodualxL$ and Binder
crossings, Figs.~\ref{fig:rhodualQ001xL_K40} and \ref{fig:BM_K40},
so there are no uncertainities where the $\Upsilon \cdot L$
crossings might flow.
We will describe overall trends along the phase boundary in the
model II in Sec.~\ref{subsec:nccp0_nccp1wAF_trends}.

Finally, from the magnetization scaling, we also estimate the exponent
$\eta$ in the two NCCP$^1$ models at $\K=0.4$, finding it to be around
$0.4 - 0.5$ in both cases on our system sizes.

\subsection {NCCP$^0$ model and NCCP$^1$ model II at $\J \to \infty$}
\label{subsec:Jinfty}

Before describing overall trends along the Photon - Higgs phase
boundary in the three models, we present Monte Carlo results
in the two $\J=\infty$ models,
Eqs.~(\ref{S_extrmnccp0}) and (\ref{S_extrmnccp1}).
In our simulations, we also use geometrical worm
updates\cite{Prokofiev, Alet} of the discrete-valued conserved
fluxes $B$.  This eliminates the critical slowing down in the NCCP$^0$
case and significantly improves statistics in the NCCP$^1$ case.
In the latter, we still use local updates of the matter fields,
which is probably why the critical slowing down persists.

As we have already described, the NCCP$^0$ model at $\J=\infty$
is equivalent to the well-studied loop model with short-range
interactions.  We use it to develop experience with the
$\rhodualqmin \cdot L$ to characterize the transition,
and we find that this measure works well.
The study also allows us to connect with the points measured along
the phase boundary in the NCCP$^0$ model at finite $\J$,
Fig.~\ref{fig:phased}a, gaining experience with finite size effects
in such matter-gauge simulations -- we will use this when discussing
overall trends in Sec.~\ref{subsec:nccp0_nccp1wAF_trends}.
Furthermore, we can now compare the two $\J=\infty$ models
Eqs.~(\ref{S_extrmnccp0}) and (\ref{S_extrmnccp1}) directly.

\begin{figure}
\centerline{\includegraphics[width=\columnwidth]{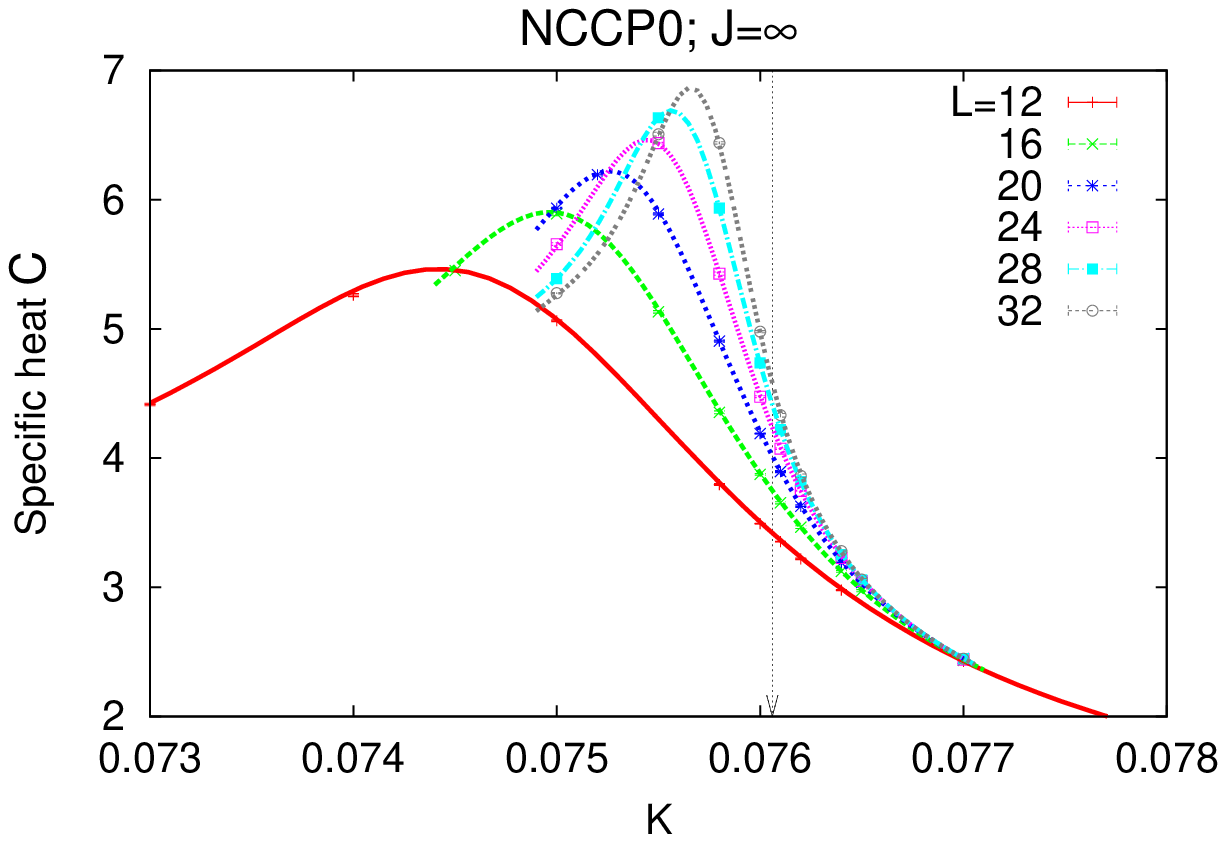}}
\centerline{\includegraphics[width=\columnwidth]{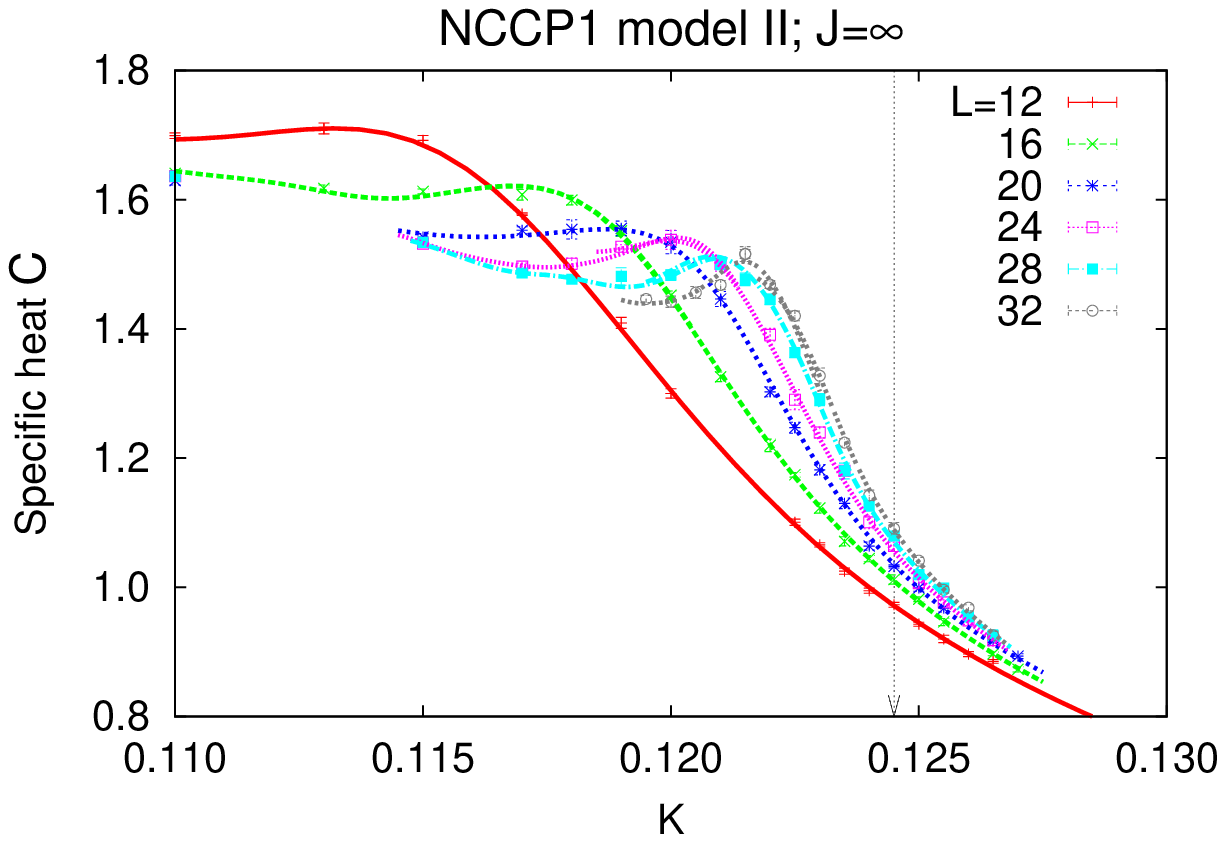}}
\caption{(color online).
Top panel: Specific heat in the NCCP$^0$ model at $\J=\infty$,
Eq.~(\ref{S_extrmnccp0}).  This is the same as a loop model
with steric interactions in an ensemble with no windings.
Bottom panel: Specific heat in the NCCP$^1$ model II at $\J=\infty$,
Eq.~(\ref{S_extrmnccp1}).
In each panel, the connecting lines are obtained by multiple
histogram method; the vertical arrow shows the bulk critical point
$\K_c$ estimated from Figs.~\ref{fig:rhodualQ001xL_Jinfty}
and \ref{fig:BM_Jinfty}.
The NCCP$^1$ transition sharpens more slowly with $L$ than NCCP$^0$
and is characterized by a larger exponent $\nu$.
}
\label{fig:C_Jinfty}
\end{figure}

In Fig.~\ref{fig:C_Jinfty}, we show the specific heat $C$ in the two
models.  In the NCCP$^0$ case, there is a clear peak which moves
towards the bulk critical point and sharpens slowly with increasing
system size.  
Analyzing the thermal singularity as in
Sec.~\ref{subsec:K40} and as described in Eqs.~(\ref{C})-(\ref{C3}), 
we obtain an estimate of the correlation length $\nu = 0.66 - 0.67$, 
which is in agreement with the accepted value $\nu=0.67$.
Turning to the NCCP$^1$ case, the specific heat signature of the
transition is weaker and appears as a feature that develops on top
of a sloping down background. The evolution of the feature with $L$
is more slow than in the NCCP$^0$ case, implying a larger exponent
$\nu$. From the study of the third cumulant of the action, we
estimate $\nu \simeq 0.75 - 0.8$.

\begin{figure}
\centerline{\includegraphics[width=\columnwidth]{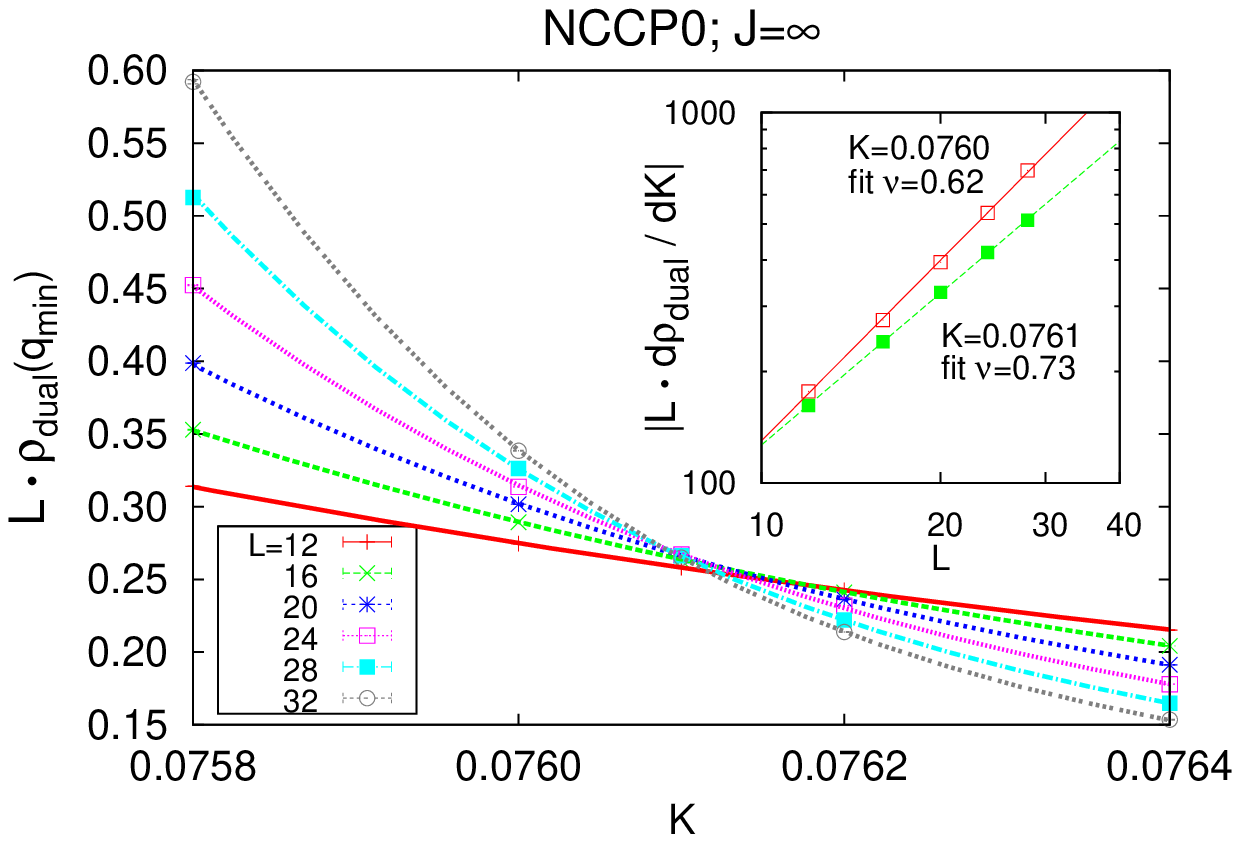}}
\centerline{\includegraphics[width=\columnwidth]{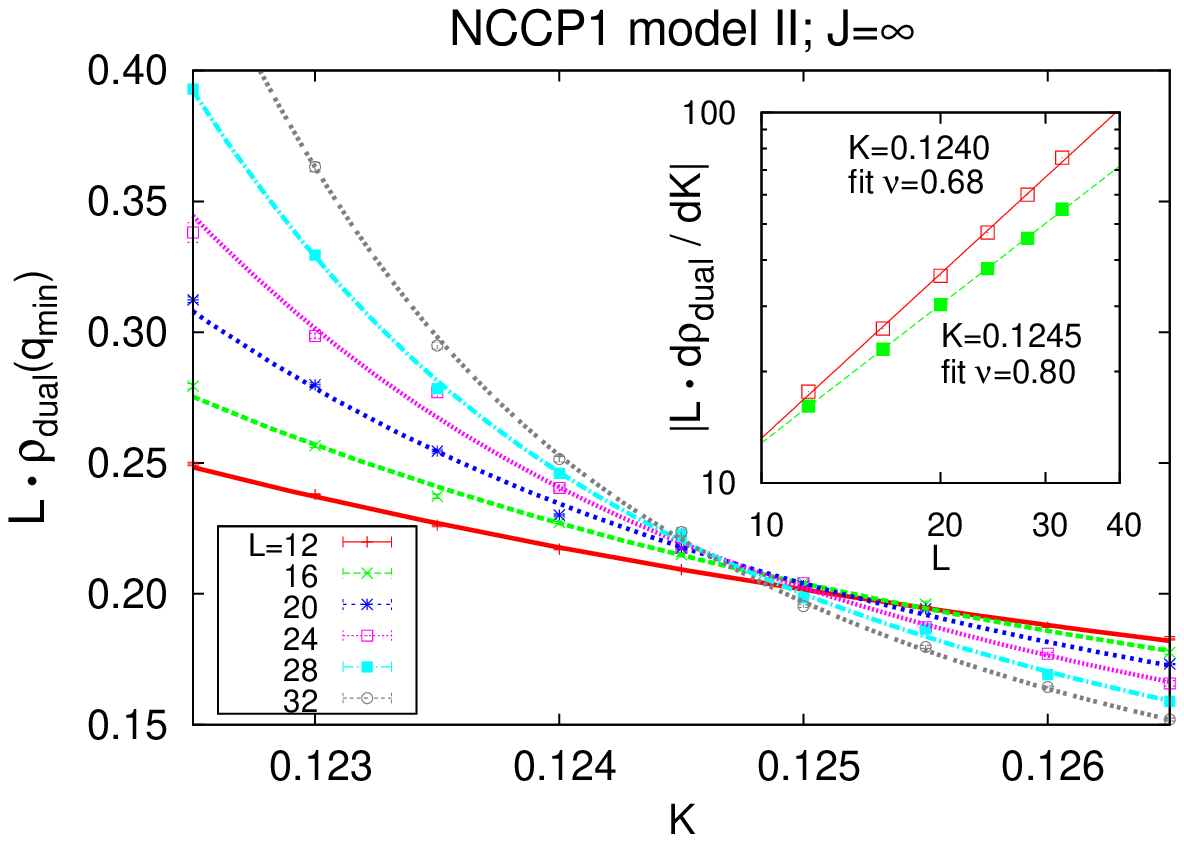}}
\vskip -2mm
\caption{(color online).
Top panel: $\rhodualqmin \cdot L$ in the NCCP$^0$ model at $\J=\infty$.
The critical point is known accurately to be $\K_c = 0.07606$
($\nu=0.67$), and the crossings converge rapidly to this location.
Inset shows the derivative of $\rhodual \cdot L$ with respect to $\K$
as a function of $L$ at two test points that bound the critical region;
the lines are direct fits to the form $\sim L^{1/\nu}$,
and the corresponding estimates of $\nu$ are also listed.
Bottom panel: Similar analysis in the NCCP$^1$ model II at $\J=\infty$;
our best estimate for $\K_c$ is $\K_c \approx 0.1245$.
}
\label{fig:rhodualQ001xL_Jinfty}
\end{figure}

Let us now characterize the transitions using the dual stiffness
shown in Fig.~\ref{fig:rhodualQ001xL_Jinfty}.  In either case,
we see a rapid convergence of the $\rhodualqmin \cdot L$ crossings
towards the putative bulk critical point.  In the NCCP$^0$ model, the
crossing for the largest two systems $L=28$ and $32$ occurs
around $\K=0.07609$,
very close to the true $\K_c=0.07606$, and the crossing location is
still moving slightly in the expected direction.
In the inset, we show the derivative $d(\rhodual \cdot L)/d\K$
as a function of $L$ at two points $\K$ that bound the critical point;
direct fits give $\nu_{\rm fit}(\K=0.0760) = 0.62$ and
$\nu_{\rm fit}(\K=0.0761) = 0.73$, which in turn bound the
expected $\nu$.  As discussed in Secs.~\ref{subsec:method}
and \ref{subsec:K40}, the inset should be viewed only as a crude
illustration of the scaling analysis, but it is already sufficient
for our comparative study.

Proceeding with similar analysis in the NCCP$^1$ case shown in the
bottom panel in Fig.~\ref{fig:rhodualQ001xL_Jinfty},
the critical point is between $\K=0.1240$ and $0.1245$.
From the analysis in the inset, we estimate $\nu$ between
$\nu_{\rm fit}(\K=0.1240) = 0.68$ and
$\nu_{\rm fit}(\K=0.1245) = 0.80$.
More elaborate scaling procedures suggest that $\nu$ is between
$0.7$ and $0.75$.  The correlation length exponent is thus detectably
larger than in the NCCP$^0$ case.

\begin{figure}
\centerline{\includegraphics[width=\columnwidth]{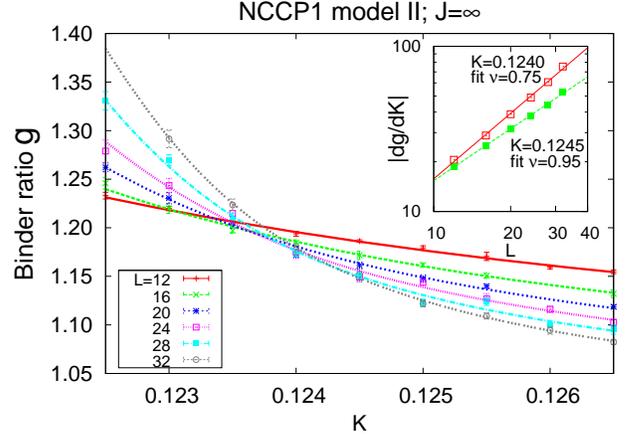}}
\vskip -2mm
\caption{(color online).
Binder ratio $g$ in the NCCP$^1$ model II at $\J=\infty$.
Comparing with the corresponding $\rhodual \cdot L$ plot in
Fig.~\ref{fig:rhodualQ001xL_Jinfty}, the $g$ crossings converge
to the same critical point bounding it from the opposite side.
Inset shows estimates of $\nu$ from the analysis of the derivative
of $g$, in the same spirit as in Fig.~\ref{fig:rhodualQ001xL_Jinfty}.
}
\label{fig:BM_Jinfty}
\end{figure}

In the NCCP$^1$ case, we also consider the Binder ratio,
Fig.~\ref{fig:BM_Jinfty}.
The crossings of $g$ bound the critical point from the opposite side
compared with the $\rhodualxL$ crossings.
Performing finite size scaling analysis of the Binder ratio curves
gives estimates of $\nu$ between 0.7 and 0.8, which is consistent with
the presented other estimates.

\begin{figure}
\centerline{\includegraphics[width=\columnwidth]{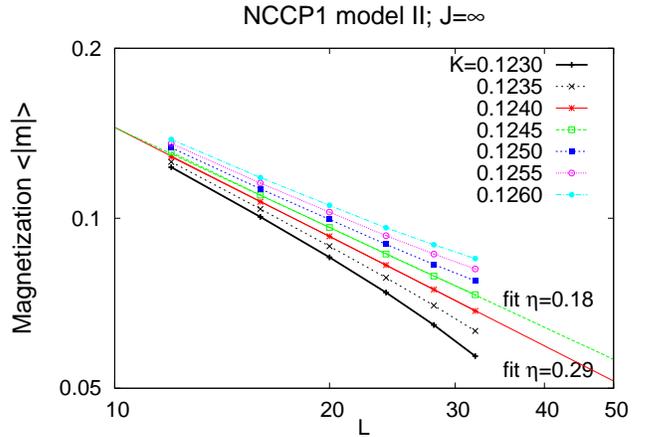}}
\vskip -2mm
\caption{(color online).
Finite size scaling analysis of the magnetization.
We expect $m \sim L^{-(1+\eta)/2}$ at criticality
and perform fits for the exponent $\eta$ at two test point.
On the log-log plot, the ordered and disordered sides curve up
and down respectively, confirming the estimates of the critical
region from Figs.~\ref{fig:rhodualQ001xL_Jinfty} and
\ref{fig:BM_Jinfty}.
}
\label{fig:m_Jinfty}
\end{figure}

Finally, in Fig.~\ref{fig:m_Jinfty}, we show finite size scaling
analysis of the magnetization.
We plot $m(\K, L)$ as a function of $L$ at fixed $\K$,
expecting Eq.~(\ref{mcrit_scaling}) at the critical point.
In the figure, there is a clear separation between the magnetically
ordered and disordered sides, and from the fits at the test points
we estimate $\eta$ between $\eta_{\rm fit}(\K=0.1240) = 0.29$ and
$\eta_{\rm fit}(\K=0.1245) = 0.18$.
Since the Binder ratios, which detect the criticality in the matter
sector, still cross near $\K=0.1240$ for our largest systems,
the correspond value of $\eta$ is likely closer to the true exponent.
When we try to scale the full magnetization curves near criticality,
we get a somewhat wider range of $\eta = 0.2 - 0.4$.

To summarize, the transition in the NCCP$^1$ model II at $\J=\infty$
has a weaker thermal singularity than in the NCCP$^0$ model;
it is a second-order transition with
$\nu_{{\rm NCCP}^1} > \nu_{{\rm NCCP}^0}$.
This conclusion is also supported by the scaling analysis of the
$\rhodualxL$ and of the Binder ratios, which give
\begin{equation}
\nu = 0.7 - 0.75, \quad ({\rm NCCP}^1 \text{ model II, } \J=\infty).
\label{nu_extrmnccp1}
\end{equation}
The magnetization order is characterized by the exponent
\begin{equation}
\eta = 0.2 - 0.4, \quad ({\rm NCCP}^1 \text{ model II, } \J=\infty).
\label{eta_extrmnccp1}
\end{equation}

\subsubsection {$\Upsilon \cdot L$ in the NCCP$^1$ model II at large
$\J = 16$}
\label{subsubsec:J160000}

\begin{figure}
\centerline{\includegraphics[width=\columnwidth]{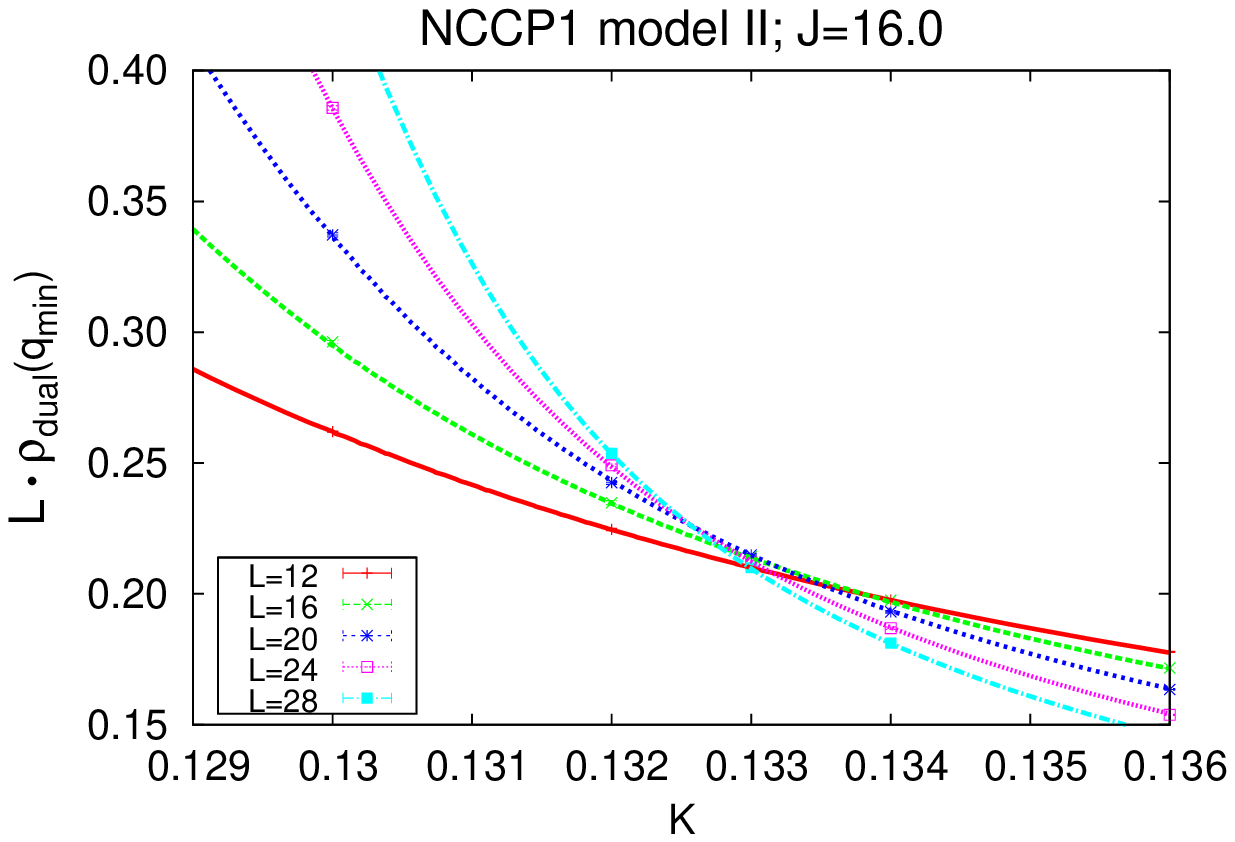}}
\vskip -2mm
\centerline{\includegraphics[width=\columnwidth]{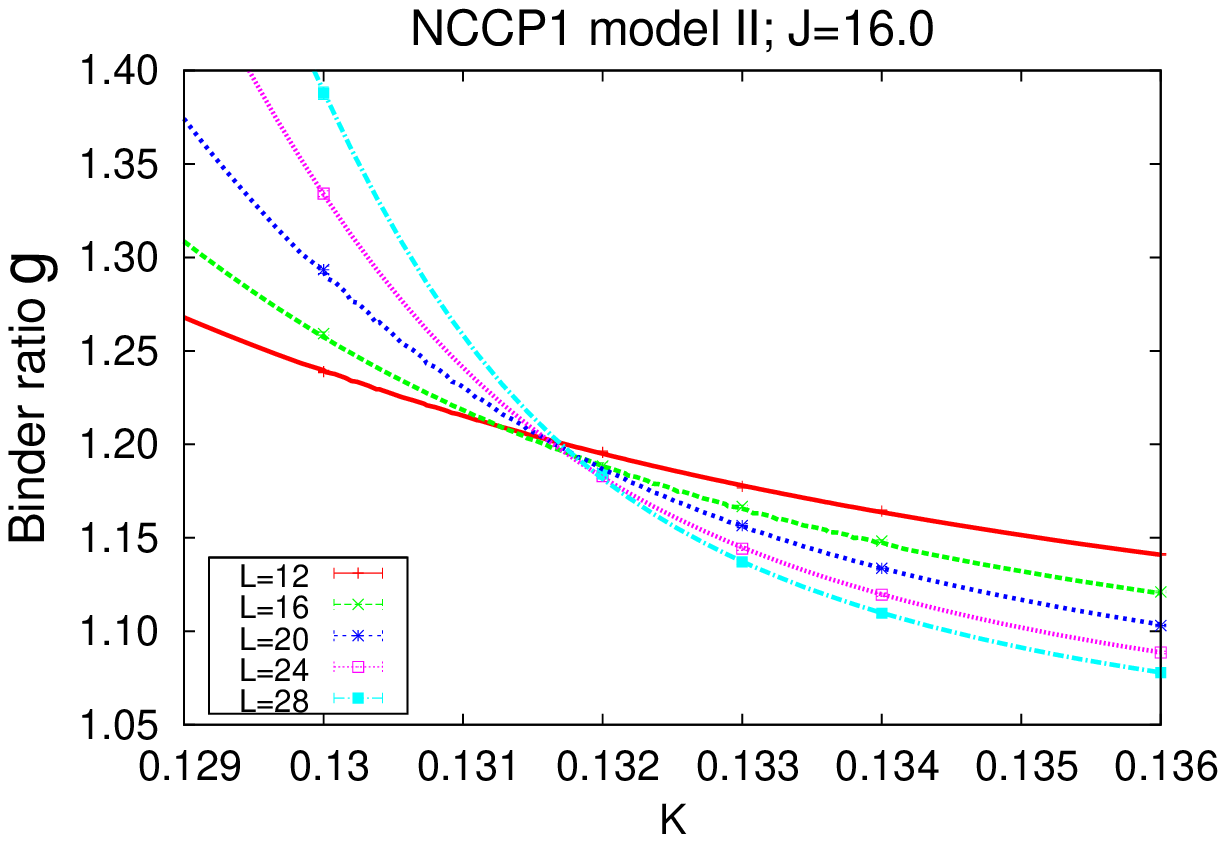}}
\vskip -2mm
\centerline{\includegraphics[width=\columnwidth]{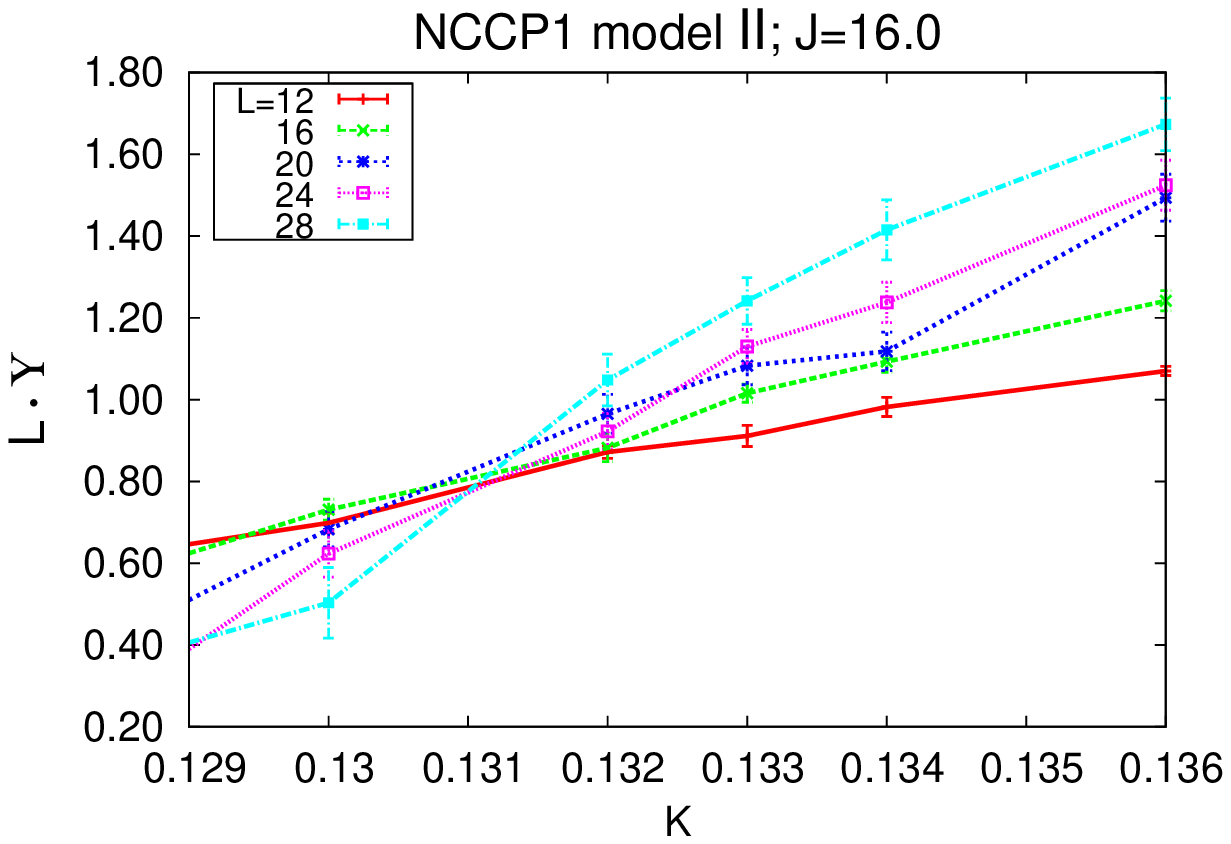}}
\vskip -2mm
\caption{(color online)
Universal crossings in the NCCP$^1$ model II at large $J=16$.
This is our closest approach towards $\J = \infty$, and the focus
here is on the boundedness of the $\Upsilon \cdot L$ at criticality.
In the top and middle panels, $\rhodualqmin \cdot L$ crossings move
towards smaller $\K$ values and the Binder ratio crossings move
towards larger $\K$ values, localizing the transition
between $\K=0.132$ and $0.1325$.
The $\Upsilon \cdot L$ data in the bottom panel is noisy, but it
is bounded by a value close to $1$ and is similar to the
$\Upsilon \cdot L$ crossings at other cuts across the phase boundary in
Fig.~\ref{fig:phased}c (compare, e.g., with the vertical cut
at $\K=0.4$ shown in Fig.~\ref{fig:rhosL_K40}).
In the top and middle panels, where the data quality is good,
the interpolating lines are obtained by the multi-histogram
reweighting technique, while in the bottom panel the lines connect
the individual data points; the error bars are estimated by
averaging over different Monte Carlo runs.
}
\label{fig:J160000}
\end{figure}

We have not been able to measure the helicity modulus $\Upsilon$
directly in the NCCP$^1$ model II at $\J=\infty$:
One expression for the $\Upsilon$ that we derived in this limit
contained singular integrals over the ${\bm z}$ variables and was
ill suited for the Monte Carlo sampling integration.
Instead, we performed a study at large but finite $\J=16$
and measured the expression Eq.~(\ref{Upsilon}).  This is our closest
connection with the $\J = \infty$ limit tracking the behavior of
$\Upsilon \cdot L$ along the phase boundary in Fig.~\ref{fig:phased}c.
The horizontal $\J=16$ cut is very far from the decoupled $\K=\infty$
limit and also far from the vertical $\K=0.4$ cut presented in
Sec.~\ref{subsec:K40}.  Nevertheless, the measured $\Upsilon \cdot L$
crossings on the two cuts are rather similar, illustrating that there is
no significant flow of the critical $\Upsilon \cdot L$ values along the
phase boundary.

The $\J=16$ study is summarized in Fig.~\ref{fig:J160000}.
The $\rhodualqmin \cdot L$ crossings and the Binder ratio crossings
behave similarly to the $\J=\infty$ model,
Figs.~\ref{fig:rhodualQ001xL_Jinfty} and \ref{fig:BM_Jinfty}.
These measures, which have smooth limits for large $\J$,
have clean statistics and allow us to accurately constrain the
critical point to lie between $\K=0.132$ and $0.1325$.
On the other hand, the $\Upsilon \cdot L$ data is much more
noisy:  Individual terms present in Eq.~(\ref{Upsilon}) can have
large typical values and large cancellations are to occur upon
the full integration.  Still, we see that the $\Upsilon \cdot L$
values near the criticality have not changed significantly from the
$\K=0.4$ cut in Fig.~\ref{fig:rhosL_K40}, despite the much larger
value of $\J$ here.

\subsection{Qualitative summary of the full Photon - Higgs phase
boundary in the NCCP$^0$ model and NCCP$^1$ model II}
\label{subsec:nccp0_nccp1wAF_trends}

We now summarize Monte Carlo results along the Photon - Higgs
phase boundary in each of the three models.
The phase diagrams were mapped out using the same criticality locators
as in the preceding sections.  Small systems up to $L = 16 - 20$
were sufficient to get good estimates of the transition points marked
in Fig.~\ref{fig:phased}, while several points were studied up to
$L = 24 - 32$ and with detail comparable to
Secs.~\ref{subsec:K40},\ref{subsec:Jinfty}.

Specifically, in the NCCP$^0$ model and in the NCCP$^1$ model II,
we studied a vertical cut at $\K=0.25$ and a horizontal cut at $\J=4$,
cf.~Figs.~\ref{fig:phased}a,c.
Combined with the study of the $\J = \infty$ limit, we thus observed
the full evolution of our measurements along the phase boundary
in the two models.
Some details will be given shortly, but we can already say
that in either model the criticality at different places along the
phase boundary is qualitatively similar to the presented $\K=0.4$ case
and connects smoothly to the $\J=\infty$ limit.
From all studies, the $\J=\infty$ models are probably closest to the
desired Higgs universalities, being least affected by other fixed
points.  In particular, these models provide our best estimates of the
critical indices.
We also studied in detail the phase boundary in the NCCP$^1$ model I,
where the situation is more complicated -- we will return to this model
in the next section.

First, we give some details in the NCCP$^0$ case.
The transitions are detected using the $\rhodualxL$ crossings
and look qualitatively similar to the $\K=0.4$ case presented
in Sec.~\ref{subsec:K40} and the $\J=\infty$ case in
Sec.~\ref{subsec:Jinfty}.
The locations of the crossings converge quickly to the critical point.
On the other hand, the specific heat peak evolves slowly with the
system size and moves towards the bulk critical point as in the top
panels of Figs.~\ref{fig:C_K40} and \ref{fig:C_Jinfty}.

One observation that we make is that the values of $\rhodualxL$
at the crossings, while apparently converged, are different at different
locations along the phase boundary -- compare, for example, the top
panels of Figs.~\ref{fig:rhodualQ001xL_K40} and
\ref{fig:rhodualQ001xL_Jinfty} -- which is at first disconcerting.
However, the observed values at the crossings approach those at
$\J=\infty$ as we move towards this point along the phase boundary.
We believe that in the thermodynamic limit, $\rhodualxL$ is universal
at criticality.  Indeed, we can show that our dual stiffness in the
one-component matter-gauge system is equal to a sum of a stiffness
$\rho$ of a mathematically equivalent discrete-valued current loop model
with short-range interactions and of a positive contribution of order
$1/(\J L^2)$.  Then, assuming universal $\rho \cdot L$ at criticality
in discrete loop models, the $\rhodualxL$ crossings in the NCCP$^0$ model
must converge to the same value but with a positive finite-size
correction of order $1/(\J L)$, which is indeed comparable to the
magnitude of the difference between the $\rhodualxL$ values in
Figs.~\ref{fig:rhodualQ001xL_K40} and \ref{fig:rhodualQ001xL_Jinfty}.
We note also that in the NCCP$^0$ model at $\K=0.4$, the initial drift
of the $\rhodualxL$ value at the crossings upon increasing our system
sizes is actually away from the $\J=\infty$ value, but we expect that
this drift will eventually turn back.  While we do not have full
understanding of all such finite-size effects in our systems,
the smooth evolution along the phase boundary and our analytical
understanding make us confident in the proposed inverted 3D XY
universality in the NCCP$^0$ model.
It is also useful to note that such corrections will have much smaller
effect on the rapid convergence of the locations of the crossings to
the bulk critical point, which is indeed what we observe.

Concluding with the NCCP$^0$ case, the scaling analysis of the dual
stiffness as in Sec.~\ref{subsec:K40} yields effective $\nu$ that
starts around 0.5-0.55 for $\K=0.4$ and moves towards the expected
value $\nu=0.67$ as we approach $\J = \infty$,
adding to our confidence that we understand this system well.

Let us now provide similar details for the NCCP$^1$ model II,
whose phase diagram has the same topology as in the NCCP$^0$ case.
Here again we see no qualitative difference in the observed criticality
as we move along the phase boundary from the $\K=0.4$ point to the
$\J=\infty$.  For example, the specific heat peaks evolve slowly to the
bulk critical points as in Figs.~\ref{fig:C_K40} and \ref{fig:C_Jinfty}.
The peak heights $C_{\rm max}$ in the vertical cut at $\K=0.25$ and in
the horizontal cut at $\J=4.0$ are essentially size-independent, i.e.,
they look even less sharp compared to the $\K=0.4$ case.
Using $\rhodualxL$ to detect the criticality, the picture that we
observe is also very similar to the NCCP$^0$ model.  In particular,
since the two systems are simulated under similar finite size conditions,
it is likely that similar effects are responsible for the apparent
difference in the $\rhodualxL$ values at the crossings as we move along
the phase boundary.
In this respect, the NCCP$^0$ model, which we otherwise understand well,
provides a useful reference on such effects that we still don't have
much experience with.  This gives us confidence when interpreting
the NCCP$^1$ case, which is our main focus.

In the NCCP$^1$ model, we also observe the transition using the
ordering of $\vec{n}$, which monitors the matter sector.
In a broad view of the phase diagram, Fig.~\ref{fig:phased}c,
for large $\K$ the Binder ratio crossings and the $\rhodualxL$
crossings start far apart for our system sizes, but as we move along
the phase boundary towards smaller $\K$, the separation in the apparent
criticality in the two sectors decreases significantly.
Recall that the gauge sector is absent in the $\K=\infty$ limit.
The convergence of such locators of the criticality in both sectors
shows that the system is sufficiently away from the $\K=\infty$ limit
and we are indeed probing the true Higgs universality.

At points between $\K=0.4$ and $\J=\infty$, scaling analyses of both the
$\rhodualxL$ and the Binder ratio give us consistently $\nu$ in the
range 0.7 to 0.75, supporting the conclusion that we are dealing with
the same continuous transition for all points on the phase boundary
in this model.
Finally, our study of the helicity modulus also agrees well with the
$\rhodualxL$ and Binder ratio measurements.
Thus, the $\Upsilon \cdot L$ crossings agree with the other locators
of the criticality.  Furthermore, everywhere on the phase boundary,
we find roughly similar values $\Upsilon \cdot L \approx 1$ at
criticality, including the cut at $J=16$ presented in
Sec.~\ref{subsubsec:J160000}, which is the largest $\J$ for which
the $\Upsilon$ was measured.  There is no significant flow of the
critical $\Upsilon \cdot L$ values along the phase boundary in
the NCCP$^1$ model II, consistent with the same continuous
transition everywhere as implied by all other measures.

\subsection{Qualitative summary of the full Photon - Higgs phase
boundary in the NCCP$^1$ model I}
\label{subsec:nccp1_trends}

\begin{figure}
\centerline{\includegraphics[width=\columnwidth]{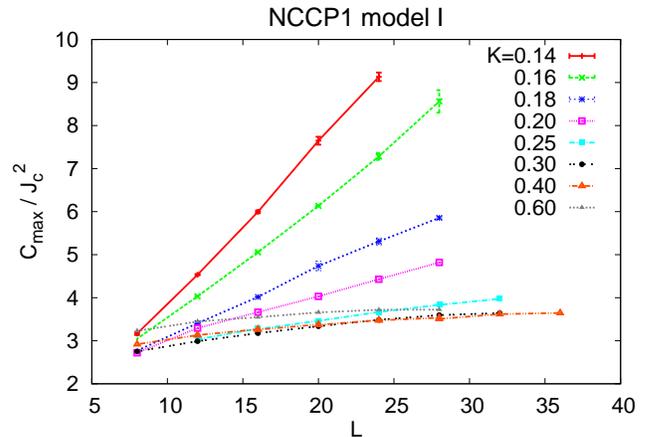}}
\caption{(color online).
Peak height of the specific heat in the NCCP$^1$ model I measured
at different locations along the Photon - Higgs phase boundary,
cf.~Fig.~\ref{fig:phased}b.  The data at $\K=0.4$ is the same as
in the corresponding inset in Fig.~\ref{fig:C_K40}.  We remove the
overall scale of the action by plotting  $C_{\rm max}/\J_c^2$,
where $\J_c$ is the critical value for a given $\K$.
Similar plot in the NCCP$^1$ model II (not shown) has $C_{\rm max}$
that depends very weakly on $L$ for all points along the phase boundary
in~Fig.~\ref{fig:phased}c.
}
\label{fig:Cmax_nccp1_all}
\end{figure}

Let us now discuss the NCCP$^1$ model I.
Fig.~\ref{fig:Cmax_nccp1_all} summarizes our measurements of the
thermal properties along the Photon - Higgs boundary,
cf.~Fig.~\ref{fig:phased}b.
For $\K \geq 0.25$, the specific heat peak height $C_{\rm max}$ grows
weakly with $L$ -- at most sublinearly for our system sizes.
For $\K = 0.18 - 0.20$, the growth is linear, and becomes faster
and dramatic for smaller $\K$.
Note that in Fig.~\ref{fig:Cmax_nccp1_all} we have divided
$C_{\rm max}$ by the corresponding $\J_c^2$ thus removing crudely the
overall scale, so it is reasonable to compare the magnitudes among
the different $\K$ data.
The $\rhodualxL$ crossings, Binder cumulant crossings, and
$\Upsilon \cdot L$ crossings are already well converged for these
system sizes (cf.~the $\K=0.4$ data in Sec.~\ref{subsec:K40}),
i.e., these systems are far from the $\K=\infty$ limit and represent
fully coupled matter-gauge criticalities.
Our interpretation of the thermal signatures is that the transition is
second-order for $\K > 0.2$ and is first order for $\K < 0.2$, while
$\K \approx 0.2$ is a candidate for the tricritical point separating the
two behaviors.
(It is worth repeating here that a similar $C_{\rm max}$ plot in the
model II, where there is no Molecular phase, has essentially
size-independent $C_{\rm max}$ for all points along the phase boundary.)

The transition at $\K=0.3$ appears more sharp compared to the $\K=0.4$
point presented in Sec.~\ref{subsec:K40}, and various broad-brush
scaling analyses give a smaller
$\nu_{\rm eff}(\K=0.3, L\leq 32) \approx 0.6$.
The transition appears still more sharp at $\K=0.2$ with
$\nu_{\rm eff} \approx 0.5$.
Also, the effective exponent $\eta$ for the magnetization order
parameter decreases towards zero as we approach $\K=0.2$.
For $\K =0.25, 0.3$, we observe that the extracted $\nu$ and $\eta$
tend to increase slightly when the scaling range around the
critical point is made more narrow.
We interpret the apparent exponent drift along the phase boundary to be
due to crossovers near the tentative tricritical point near $\K=0.2$.
(To repeat here, we see no such exponent drift in the model II).

Certainly, this drift is an important concern that we have been
unable to fully address in the model I with our system sizes.
These difficulties are the main reason why we also considered the
NCCP$^1$ model II, where the first order region is absent and the
systematics of the scalings is improved, as we have already described.
This experience with the model II gives us better grounds for making
the above interpretation in the model I.  It would be useful to
confront this with direct simulations in the model I on larger
system sizes than in our work.
Larger systems should allow one to better filter out the crossovers
and focus on the true critical behavior, e.g., confronting our
expectation of the universal exponents along the continuous region of
the phase boundary.

To give more details behind the proposed scenario in the NCCP$^1$
model I, we also present the behavior of the Binder ratio crossings and
$\Upsilon \cdot L$ crossings along the Photon - Higgs phase boundary.
These are vital locators of the criticality and are expected to
converge to universal values in the second-order case.
Going back to Figs.~\ref{fig:BM_K40} and \ref{fig:rhosL_K40} at $\K=0.4$,
we see some weak evolution of the values at the crossings.
We find that these values further drift as we move along the
phase boundary.  Similarly to the drifts in the apparent exponents,
this is an important concern.
In this regard, it is helpful to have a picture of all data along
the phase boundary.  We can then try to see possible attracting points
of such drifts of the crossings and also see where the presented
detailed study at $\K=0.4$ of Sec.~\ref{subsec:K40} stands in the
overall picture.

\begin{figure}
\centerline{\includegraphics[width=\columnwidth]{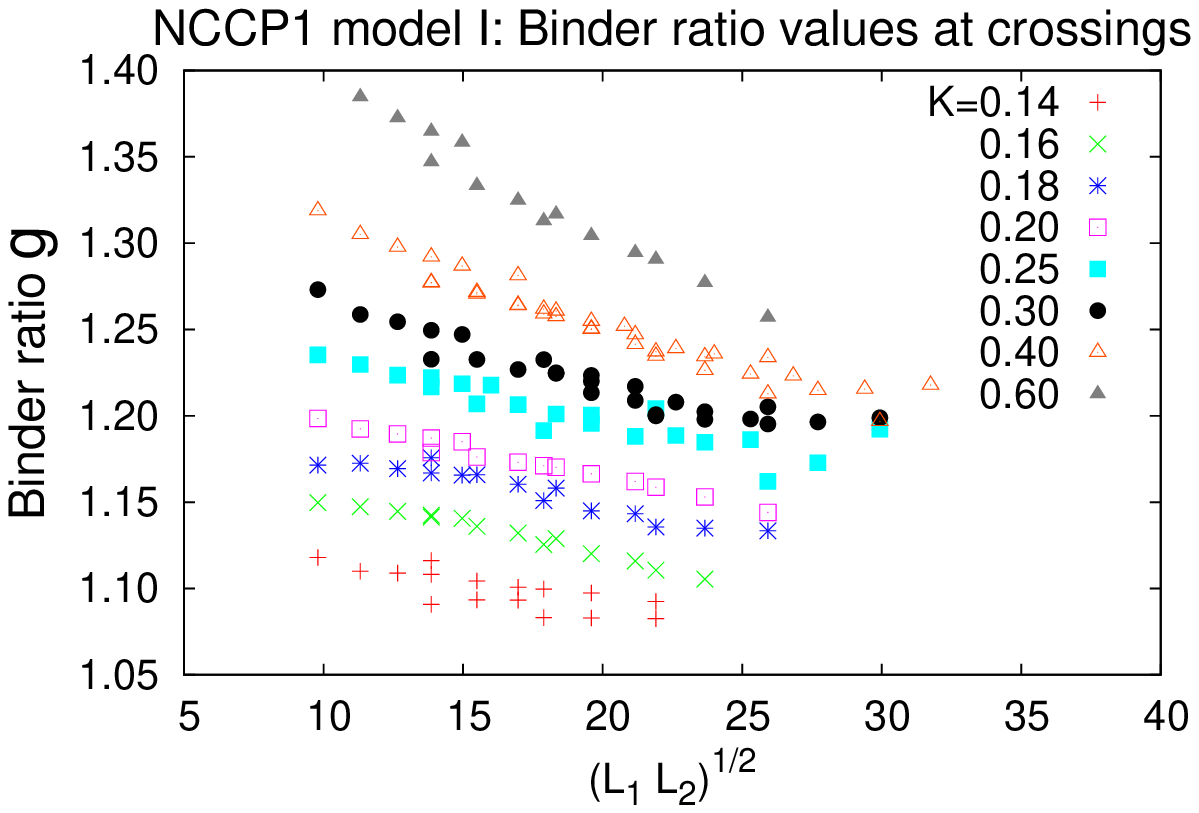}}
\centerline{\includegraphics[width=\columnwidth]{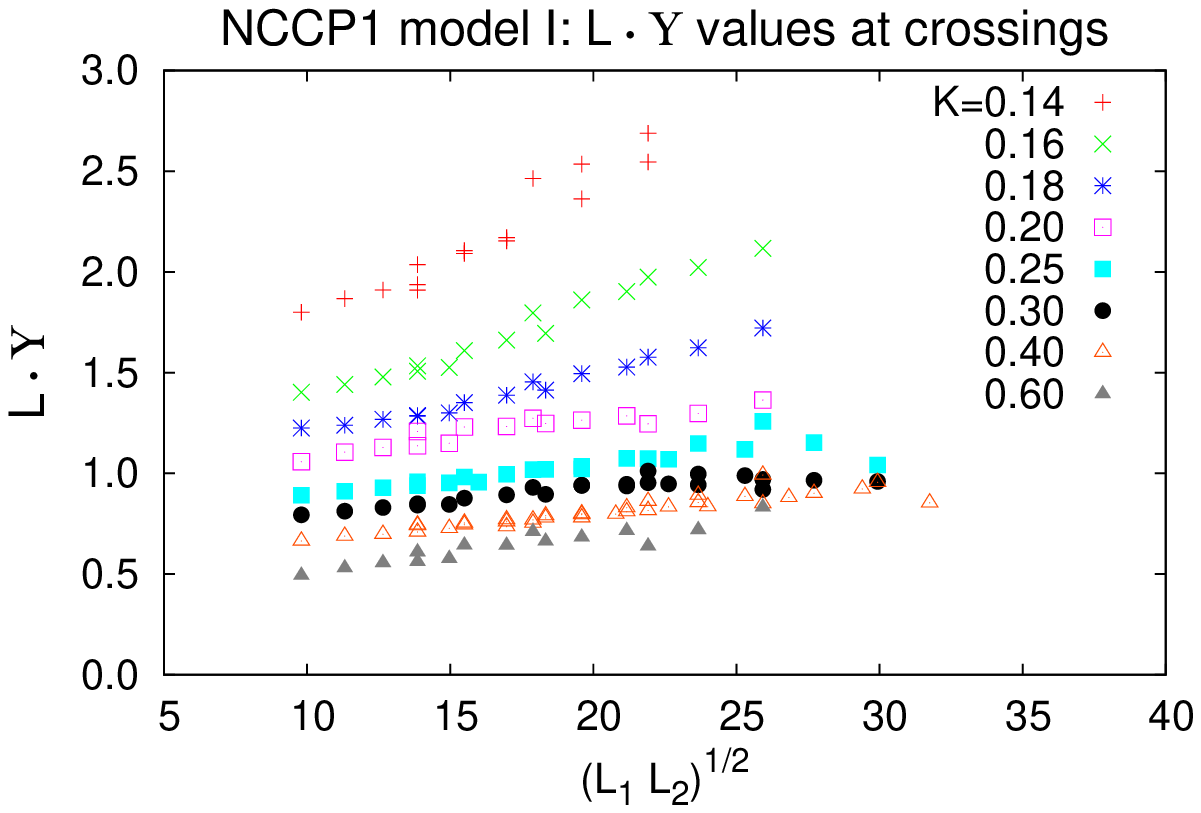}}
\vskip -2mm
\caption{(color online).
Top panel:  Binder ratio values at the corresponding crossings
for all pairs of system sizes $L_1$ and $L_2$.
The data is collected by analyzing $\J$ scans such as
Fig.~\ref{fig:BM_K40} at a number of $\K$ locations along the
Photon - Higgs phase boundary from Fig.~\ref{fig:phased}b.
For an illustration, the pairs of system sizes are organized by
$\sqrt{L_1 L_2}$ on the horizontal axis.
Bottom panel:  $\Upsilon \cdot L$ values at the corresponding
crossings obtained by analyzing the helicity modulus plots such as
Fig.~\ref{fig:rhosL_K40}.
Together with Fig.~\ref{fig:fixedBM1p15} and all other data,
our interpretation is that the transition is continuous for $\K > 0.2$
and becomes first order for $\K < 0.2$.
}
\label{fig:crossings_alldata}
\end{figure}

The top panel in Fig.~\ref{fig:crossings_alldata} summarizes all
collected data for the Binder ratio crossings.
For each $\K$, we consider a vertical $\J$ scan across the phase
boundary in Fig.~\ref{fig:phased}b and analyze the Binder ratios
as done in Fig.~\ref{fig:BM_K40}.  For each pair of system sizes
$L_1$ and $L_2$, we find the corresponding crossing of the curves
and organize the results by showing
$\{ \sqrt{L_1 L_2},~g_{\rm cross}(L_1, L_2; \K)\}$.
Admittedly, the plot Fig.~\ref{fig:crossings_alldata} is noisy,
since the crossings are fairly sensitive to the statistical errors.
For some systems we show results obtained from several independent
data samples, so the spread of the symbols gives some idea about our
uncertainties.  Generally, we do not trust the values when the
crossings become non-systematic with $L$, which happens for our
largest systems.
From the available data, we see that for $\K > 0.2$ the Binder ratio
values evolve strongly initially but then appear to saturate to a
value $g \approx 1.2$ for the largest systems.  On the other hand,
for $\K < 0.2$ the Binder ratios at the crossings flow to smaller values.

Note that in the scenario with the tricritical point, we expect
that the Binder ratios go to different nontrivial values in each of
the three cases -- at the second-order line, at the tricritical point,
and at the first-order transitions.
In this respect, the fact that the initial evolution in all cases is
in the same direction makes the interpretation of the
Binder crossings data more difficult.  Certainly, improving the
statistics of the data in Fig.~\ref{fig:crossings_alldata} and going to
larger sizes could address such concerns and confront our interpretation,
which we nevertheless make given all available data.

A more clear separation between the different regimes is provided by
similarly collected data for the $\Upsilon \cdot L$ crossings shown
in the bottom panel in Fig.~\ref{fig:crossings_alldata}.
In the second-order case, we expect the crossings to go to a universal
value, and we see that for $\K > 0.2$ these indeed appear to accumulate
to a value $\Upsilon \cdot L \approx 1$ for the largest systems.
On the other hand, in the first-order regime, we expect the
$\Upsilon \cdot L$ crossing values to increase linearly with the system
sizes, and we indeed see such growth for $\K < 0.2$.
We are fairly confident that in the proposed second-order region the
values of $\Upsilon \cdot L$ do not grow without limit at the
transition, since we can bound the possible drift of the
$\Upsilon \cdot L$ crossing locations by the $\rhodualxL$ crossing
locations that converge to the transition from the opposite side,
cf.~Figs.~\ref{fig:rhodualQ001xL_K40}, \ref{fig:BM_K40}, and
\ref{fig:rhosL_K40}.
(We repeat here that the $\Upsilon \cdot L$ crossings in the NCCP$^1$
model II were also found to be bounded and of order $1$ at several
different places along the phase boundary;
the situation in the proposed second-order regime in the NCCP$^1$
model I is consistent with this in the NCCP$^1$ model II.)

\begin{figure}
\centerline{\includegraphics[width=\columnwidth]{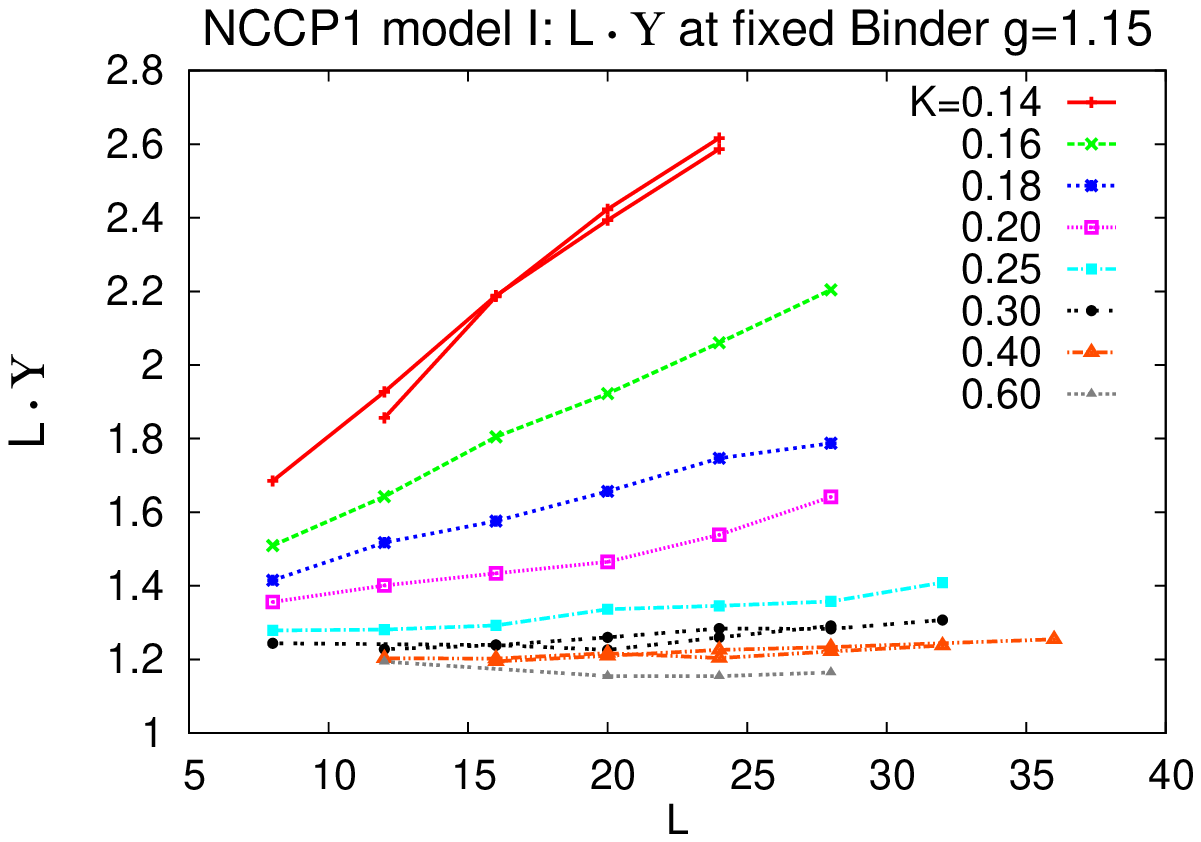}}
\centerline{\includegraphics[width=\columnwidth]{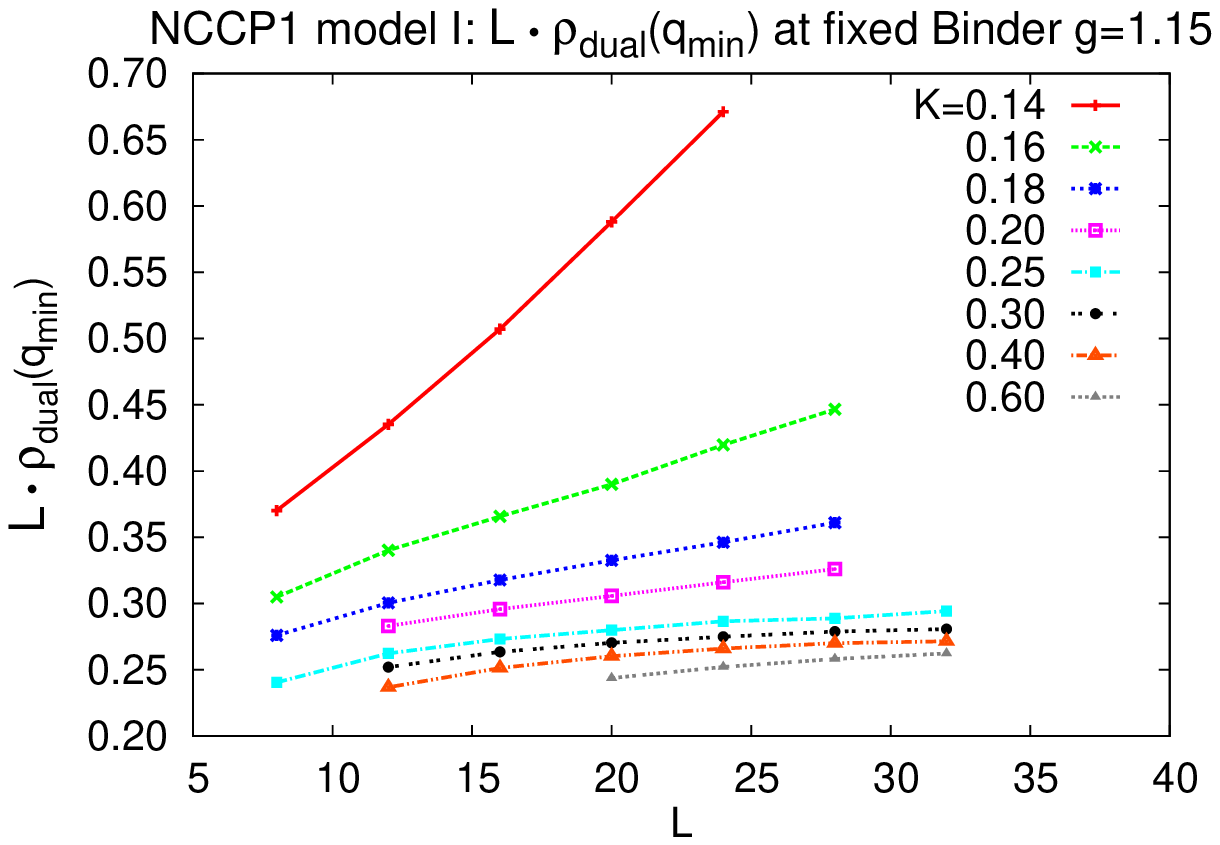}}
\vskip -2mm
\caption{(color online).
Different view of the Binder ratio, $\Upsilon \cdot L$, and
$\rhodualxL$ data.
For each system size, we define the finite-size critical point
to be where the Binder ratio takes value $g=1.15$ and then measure the
$\Upsilon \cdot L$ and $\rhodualxL$ at this point and plot the results
as a function of $L$.
Our interpretation of the data is that the transition is continuous
for $\K > 0.2$ and becomes first order for $\K < 0.2$.
}
\label{fig:fixedBM1p15}
\end{figure}

In Fig.~\ref{fig:fixedBM1p15}, we present a somewhat different
perspective on the collected Binder ratio, $\Upsilon \cdot L$, and
$\rhodualxL$ data.
For each $\K$, we define an operational critical point $\J_c(L)$ for a
given size $L$ as the point where the Binder ratio takes value $g=1.15$.
We then measure $\Upsilon \cdot L$ and $\rhodualxL$ at this point and
plot the results as a function of $L$.
The specific value $g=1.15$ lies between the apparent crossing values
in the $\K > 0.2$ and $\K < 0.2$ regimes and is selected so that
$\J_c(L)$ approaches the true critical point systematically from the
large $\J$ values when $\K > 0.2$.
Any $g$ between the limits on the two sides of the transition can
be used for such a definition of the finite-size critical point,
and we have checked that the results do not change qualitatively
if we fix the Binder ratio at different values such as
$g=1.3, 1.2, 1.1$ instead.

From the top panel in Fig.~\ref{fig:fixedBM1p15}, we see that
thus measured $\Upsilon \cdot L$ values hardly change with the
system size for $\K > 0.2$ and all are in a narrow range around
$1.2 - 1.3$.
There may be slight systematic drift, and the ultimate fate of this is
hard to judge given the uncertainties of our measurements and limited
system sizes.  However, we note that the relative variations of the
$\Upsilon \cdot L$ values are small:
Had we chosen the vertical axis in the figure to start at zero,
which is natural given that there is no other scale for the expected
universal values of $\Upsilon \cdot L$, the curves for $\K > 0.2$
would look essentially flat.  Moreover, the values are fairly close
in a whole range range $\K = 0.25 - 0.6$, and are also similar to
the values we find at this Binder ratio in the NCCP$^1$ model II
along the whole phase boundary (e.g., near points $\K=0.4, 0.25$ and
$\J=4.0, 16.0$ in Fig.~\ref{fig:phased}c).
This is to be contrasted with the behavior for $\K < 0.2$,
where we see a strong growth of the critical $\Upsilon \cdot L$
with the system size starting from the small $L$, which is interpreted
as first-order behavior.

Similar considerations apply to the $\rhodualxL$ measured at fixed
$g=1.15$ and shown in the bottom panel in
Fig.~\ref{fig:fixedBM1p15}.  For $\K < 0.2$, the $\rhodualxL$
values grow strongly with the system size, while for $\K > 0.2$
they grow very weakly and tend to saturate.  We should mention that we
do not see such clear separation between the two regimes by looking
at the pairwise crossings of $\rhodualxL$ (not shown):
The values at the crossings are essentially size-independent for
all $\K$ for our sizes, i.e., we do not see the first order region
in this measure yet.
We also want to point out that from the study of the model II at
$\J=\infty$, Fig.~\ref{fig:rhodualQ001xL_Jinfty},
we expect the critical value of $\rhodualxL$ to be eventually smaller
than what we see at the crossings in the model I.
That is, there are strong finite size corrections to this measure,
but such tendencies are similar to what we have observed in the
NCCP$^0$ model and NCCP$^1$ model II and discussed earlier.

We have also analyzed the evolution of the extracted $\eta$ exponent
in the two NCCP$^1$ models along the phase boundary.
In the model II, we see a convergence with the system size towards the
value $\eta \approx 0.3$ estimated in the $\J \to \infty$ limit,
cf.~Eq.~(\ref{eta_extrmnccp1}).
On the other hand, the effective $\eta$ is still larger around 0.4-0.5
in the model I at $\K=0.6, 0.4$ for our system sizes but is decreasing.
The initial large effective $\eta$ is expected, given that this
`Molecular' order parameter has $\eta \approx 1.37$ at the O(4)
point\cite{Isakov} where the matter and gauge fields are decoupled.
This also explains why the original small system study of the
NCCP$^1$ model I at $\K=0.6$ in Ref.~\onlinecite{shortlight}
obtained large $\eta$, which we now revise.

To conclude, despite some uncertainties with the weak drifts as
we move along the Photon - Higgs phase boundary in the NCCP$^1$ model I,
we believe that there is a wide regime where the transition is
continuous.  While the discussion of the full phase boundary is
sketchy (in particular, we have not tried to accurately characterize
the tricritical and first-order regimes), it provides better
understanding of possible issues and concerns in our numerical
explorations of the Higgs transition, and, in particular, gives
a better idea where the detailed study at $\K = 0.4$ of
Sec.~\ref{subsec:K40} stands in the overall picture.
We conclude that $\K = 0.4$ is well inside the continuous regime;
recall that it is also far from the $\K = \infty$ limit, so we have
strong reasons to claim that it is representative of the true
continuous Higgs criticality.

\section{Summary and conclusions}
\label{sec:concl}

To summarize, we perform a comparative Monte Carlo study of the
Higgs transition in the one-component and two-component lattice
superconductors and argue for the generic second-order nature in
both cases.  The one-component model (called NCCP$^0$ here),
which is well understood analytically and numerically, serves as a
reference system for our matter-gauge simulations.

We first examine in detail the two-component model introduced in
Ref.~\cite{shortlight} and called NCCP$^1$ model I here.
The phase diagram is shown in Fig.~\ref{fig:phased}b, and we argue
that there is a range of parameters, $\K > 0.2$, where the transition
is second-order.  The main evidence is the weakness of the thermal
singularities, but other measures are also in favor of the second-order
nature.  We present an example of such an analysis at $\K=0.4$ and
argue that the correlation length exponent is larger than in the
NCCP$^0$ case.

The NCCP$^1$ model I also has the Molecular phase.  In the vicinity
where the three phases meet, the thermal signatures of the Photon to
Higgs transition change dramatically, see Fig.~\ref{fig:Cmax_nccp1_all}.
We interpret this as the transition becoming first order for
$\K < 0.2$, which is also expected on general grounds due to increasing
fluctuations of the molecular field, see Appendix~\ref{app:mf}.
This interpretation implies a tricritical point around
$\K \approx 0.2$ separating the two regimes.

It is difficult to characterize firmly the region of continuous
transitions in the model I.  At each point, scaling apparently works,
but the extracted exponents drift as we move along the phase boundary,
with both $\nu$ and $\eta$ decreasing towards $0.5$ and $0$ respectively
as we approach $\K=0.2$.  We attribute this to crossovers near the
tricritical point.  In this picture, the second-order region is bordered
by the O(4) transition at $\K=\infty$ and the tricritical point near
$\K=0.2$.  The RG flows near the corresponding fixed points are causing
a wealth of crossovers.  The absence of full leverage over such
phenomena is the main source of our difficulties in this model.
It is worth repeating that in this picture the Higgs transition is
controlled by a new fixed point, and there is no contradiction here
since the RG flows are likely occurring in a more complex space than
just the $\K - \J$ parameter space of the specific model.
By studying the dual stiffness and comparing with the NCCP$^0$ model,
we know that our systems are sufficiently away from the $\K=\infty$
limit, and the apparently continuous transition with large $\nu$ is
most naturally explained as the second-order Higgs universality.

To alleviate the difficulties associated with the crossovers
near the putative tricritical point, we also consider a modification
of the original NCCP$^1$ model by adding short-range antiferromagnetic
interactions that hinder the formation of the Molecular phase.
In fact, in the specific NCCP$^1$ model II, the Molecular phase is
eliminated completely, allowing one to focus solely on the Photon to
Higgs transition.  We find that the transition is continuous along the
whole phase boundary, see Fig.~\ref{fig:phased}c.
The scalings are consistent throughout, giving the large correlation
length exponent.  The transition line connects smoothly with the
$\J=\infty$ limit, which can be viewed as a description in terms of the
Abrikosov-Nielsen-Olesen vortices that have nontrivial internal
structure.
We are able to realize this structure and study the transition in
Monte Carlo using local matter degrees of freedom, and this numerics
gives our best estimates of the critical indices $\nu$ and $\eta$,
Eqs.~(\ref{nu_extrmnccp1}) and (\ref{eta_extrmnccp1}).
For the future, it would be useful if one could realize such a
nontrivial loop system in a manner that would eliminate critical
slowing down, as is possible for the ANO vortices of the one-component
model, which are short-range interacting loops with no internal
structure.

Let us conclude with more questions that one would like to study.
First of all, we mention possible applications to unusual quantum
critical phenomena in magnets.\cite{deccp, SU2ring1, SU2ring2}
Our correlation length and magnetization exponents are broadly in 
agreement with those found in the study of the continuous Neel to
Valence Bond Solid transition in an SU(2) symmetric spin-1/2 system
in Refs.~\onlinecite{SU2ring1, SU2ring2}.
(It would also be useful to compare universal amplitudes
as discussed in Ref.~\onlinecite{Kaul}.)
Motivated by this application, it would be useful to find the scaling
dimension of a monopole insertion operator in the gauge theory
Eq.~(\ref{S_continuum}), since this corresponds\cite{deccp} to the
VBS order parameter measured in Ref.~\cite{SU2ring1, SU2ring2}.
Our $\J=\infty$ formulation of the model II, where inserting a
monopole (antimonopole) corresponds to a source (sink) for the discrete
$B$-fluxes, may be well suited for this.  It would also be useful
to find the scaling dimension of multiple monopole insertions.
For example, one would like to check whether quadrupled monopoles are
irrelevant as proposed in the deconfined criticality scenario for the
Neel - VBS transition on the square lattice and as needed for the
non-compact gauge theory Eq.~(\ref{S_continuum}) to be applicable to
this transition.\cite{deccp}

It would be also useful to know the effect of various deformations
away from the SU(2)-invariant case.
For example, recently Alet\etal\cite{dimers3D} studied a Coulomb to
Valence Bond Solid transition in a classical dimer model on a 3D cubic
lattice.  This can be (approximately) mapped\cite{Bergman} to
Eq.~(\ref{S_continuum}) plus cubic anisotropy terms, which, however,
appear only with more derivatives/fields.
Alet\etal\cite{dimers3D} found a continuous transition but with
different critical indices from ours.  Several possible explanations are
given in Ref.~\onlinecite{dimers3D}, and more studies would be
worthwhile.

Finally, we would also like to revisit the U(1)$\times$U(1) variant
using the insights learned here.

\acknowledgments 

We have benefited from useful discussions with
M.~P.~A.~Fisher, D.~Huse, T.~Senthil, and A.~B.~Kuklov.  
The numerical work was started at KITP, and use of the Hewlett-Packard
and CNSI Computer Facilities at UCSB is gratefully acknowledged. 
The numerical work at Caltech was performed using the IT2 computer
system operated by the Caltech CACR.  
A.V. acknowledges support from NSF DMR-0645691 and 
O.I.M. from the A.~P.~Sloan Foundation.


\appendix

\section{Mean field analysis near the Molecular phase}
\label{app:mf}

\begin{figure}
\centerline{\includegraphics[width=\columnwidth]{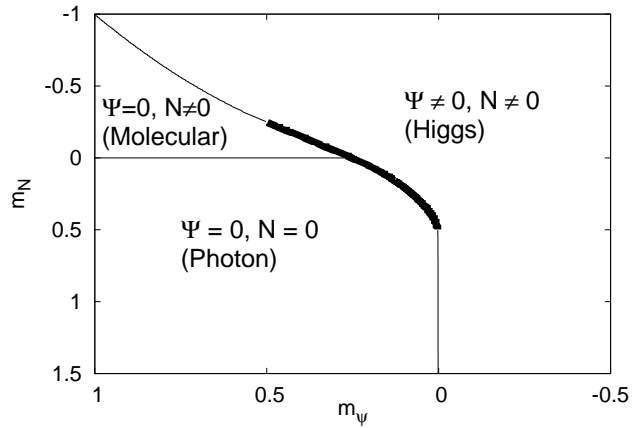}}
\vskip -2mm
\caption{Phase diagram of the Landau mean field theory
Eq.~(\ref{Smf}) in the vicinity where the three phases meet.
The bold lines mark where the transitions become first order
in the vicinity of the fork.  We chose the axes so that
$m_\psi, m_N > 0$ in the lower left corner, which gives the
locations of the phases roughly similar to that in
Fig.~\ref{fig:phased}b.
}
\label{fig:mfphased}
\end{figure}

Here we summarize the mean field argument that in the two-component
system, the Photon to Higgs transition becomes first order in the
vicinity of the Molecular phase.\cite{Kuklov}
We introduce an order parameter $\vec{N}$ for the molecular phase
and write a schematic Landau theory that contains all three phases:
\begin{eqnarray}
S[{\bm \Psi}, \vec{N}] &=&
m_\psi |{\bm \Psi}|^2 + \frac{u_\psi}{2} |{\bm \Psi}|^4
+ m_N \vec{N}^2 + \frac{u_N}{2} (\vec{N}^2)^2 \nonumber \\
&-& g \vec{N} \cdot {\bm \Psi}^\dagger \vec{\sigma} {\bm \Psi} ~.
\label{Smf}
\end{eqnarray}
Setting $u_\psi = u_N = g = 1$ and minimizing this action, we obtain
the phase diagram in Fig.~\ref{fig:mfphased}. Away from the fork all
transitions are continuous, while near the fork the Photon to Higgs
and Molecular to Higgs transitions become first order. Let us see
how this happens for the Photon to Higgs transition. For large
positive $m_N$, we can ``integrate out'' the $\vec{N}$ field and
obtain the renormalization $u_\psi \to u_\psi - g^2/(2 m_N)$. When
we approach the Molecular phase, $m_N$ becomes smaller leading to a
sign change for the $|\Psi|^4$ term. By the familiar mechanism in
the Landau theory, the transition becomes first order due to such
short-distance energetics near the molecular phase.

\section{Evaluation of the Helicity modulus $\Upsilon$ in the
NCCP$^1$ models}
\label{app:helicity}
The helicity modulus is defined from the dependence of the free energy
on the twist in the boundary conditions.
As is common when calculating stiffnesses in Monte Carlo,
the twist angle can be distributed into small twists throughout the
system, and the helicity modulus can be then obtained by measuring
appropriate derivative expressions during simulations with periodic
boundary conditions.
In this approach, the expressions to be evaluated depend on the
detailed interactions and are different in the NCCP$^1$ model I and
model II cases.

Specifically, we consider twisting the vector $\n = (n_1, n_2, n_3)$
in the $n_1 - n_2$ plane, which is achieved by the substitution
$(z_{r\up}, z_{r\dn}) \to
 (z_{r\up} e^{i \gamma r_\mu}, z_{r\dn} e^{-i \gamma r_\mu})$
in the action for the twist imposed in the lattice direction $\hat\mu$.
Carrying through the derivations, we can summarize the expressions
to be measured in the two models as
\begin{widetext}
\begin{eqnarray}
\Upsilon_\mu L^d &=&
\sum_i \J
\left\langle
{\rm Re}[{\bm z}_i^\dagger {\bm z}_{i+\hat\mu} e^{i a_{i\mu}}]
\right\rangle \nonumber \\
&-& \sum_i 4 \J^2
\left\langle
  f(\J |{\bm z}_i^\dagger {\bm z}_{i+\hat\mu}|) \,
  \text{Re}[z_{i\up}^* z_{i\dn} z_{i+\hat\mu,\dn}^* z_{i+\hat\mu,\up}]
\right\rangle
+ \sum_i 4 \J^4
\left\langle
  g(\J |{\bm z}_i^\dagger {\bm z}_{i+\hat\mu}|) \,
  \left({\rm Im}[z_{i\up}^* z_{i\dn} z_{i+\hat\mu,\dn}^* z_{i+\hat\mu,\up}]
  \right)^2
\right\rangle
\nonumber \\
&-&
\left\langle
\left(
\sum_i \J\, {\rm Im}[z_{i\up}^* z_{i+\hat\mu,\up} e^{i a_{i\mu}}
                   - z_{i\dn}^* z_{i+\hat\mu,\dn} e^{i a_{i\mu}}]
- \sum_i 2 \J^2\, f(\J |{\bm z}_i^\dagger {\bm z}_{i+\hat\mu}|) \,
      {\rm Im}[z_{i\up}^* z_{i\dn} z_{i+\hat\mu,\dn}^* z_{i+\hat\mu,\up}]
\right)^2
\right\rangle ~,
\label{Upsilon}
\end{eqnarray}
\end{widetext}
where we introduced two functions $f(x)$ and $g(x)$.
In the model I case, $f(x) = g(x) = 0$.
In the model II case, they are
\begin{eqnarray}
f(x) = \frac{I_1(x)}{x I_0(x)} ~, \quad\quad
g(x) = \frac{f^\prime(x)}{x} ~,
\end{eqnarray}
where $I_0(x)$ and $I_1(x)$ are the modified Bessel functions;
the corresponding terms in the expression for $\Upsilon$ arise from the
additional interactions defining the model II, Eq.~(\ref{nccp1wAF}).



\begin{thebibliography}{x}
\bibitem{derivs_footnote}
The derivatives in the insets in
Figs.~\ref{fig:rhodualQ001xL_K40}, \ref{fig:BM_K40}, and
\ref{fig:rhosL_K40} are calculated in the Monte Carlo process.
In the NCCP$^0$ model and the NCCP$^1$ model I, the derivative is with
respect to the parameter $\J$ that multiplies the link energy,
while in the NCCP$^1$ model II the derivative is with respect to a
formal parameter $\beta$ that would multiply the full energy
(this derivative is less costly to compute and such different choices
have little effect on the extracted exponents).  We have not calculated
the derivative of the helicity modulus $\Upsilon$ in the NCCP$^1$ model II
because of the complexity of the expression Eq.~(\ref{Upsilon})
for the $\Upsilon$ itself.
\end{thebibliography}
\end{document}